\documentclass[11pt,a4paper]{article}
\usepackage[utf8]{inputenc}
\usepackage{amsmath}
\usepackage{placeins}
\usepackage{url}
\usepackage{natbib}
\usepackage{color,xcolor,setspace,geometry}
\setcitestyle{authoryear,open={(},close={)}}
\usepackage{cite}
\usepackage{amsfonts}
\usepackage[normalem]{ulem}
\usepackage{pifont}

\usepackage{arydshln}

\usepackage{tikz}
\usetikzlibrary{shapes}
\usetikzlibrary{plotmarks}
\usetikzlibrary{tikzmark,matrix,calc}
\usetikzlibrary{arrows.meta,calc,decorations.markings,math,arrows.meta}

\usepackage{amssymb}
\usepackage{multirow}
\usepackage{graphicx}
\usepackage{soul}
\soulregister\citep7
\soulregister\cite7
\soulregister\citet7
\soulregister\citealp7
\soulregister\ref7
\usepackage{rotating}
\usepackage{setspace}
\usepackage{threeparttable}
\usepackage{authblk}
\usepackage{subfigure}

\usepackage{appendix}

\begin{document}
\begin{spacing}{1.5}	

\title{Subscription-Based Inventory Planning for E-Grocery Retailing}

\author[$\dag$ $\star$]{David Winkelmann}
\author[$\ddag$]{Charlotte Köhler}

\affil[$\dag$]{Department of Business Administration and Economics \\ Bielefeld University, Bielefeld, Germany}
\affil[$\ddag$]{Department of Data Science \& Decision Support\\ European University Viadrina, Frankfurt (Oder), Germany}
\affil[$\star$]{Corresponding author: david.winkelmann@uni-bielefeld.de}

\maketitle
\noindent

\begin{abstract}
The growing e-grocery sector faces challenges in becoming profitable due to heightened customer expectations and logistical complexities. This paper addresses the impact of uncertainty in customer demand on inventory planning for online grocery retailers. Given the perishable nature of grocery products and intense market competition, retailers must ensure product availability while minimising overstocking costs. We propose introducing subscription offers as a solution to mitigate these inventory challenges. Unlike existing literature focusing on uniform subscription models that may harm profitability, our approach considers the synergy between implementing product subscriptions and cost savings from improved inventory planning. We present a three-step procedure enabling retailers to understand uncertainty costs, quantify the value of gathering additional planning information, and implement profitability-enhancing subscription offers. This holistic approach ensures the development of sustainable subscription models in the e-grocery domain.

\end{abstract}

\section{Introduction}
In the realm of e-grocery retailing, customers place online orders for delivery at a designated date and time, commonly referred to as \textit{attended home delivery}. Despite the substantial growth in the e-grocery sector in recent years (see, e.g.\ \citealp{hubner2019distribution}), many retailers have struggled to achieve profitability, leading some to exit the market. This can be attributed to the elevated customer expectations and the logistical complexities faced by retailers, which contribute to high operational costs and hinder the attainment of profitability. The process of attended home deliveries can be divided into three key steps, as outlined by \citet{campbell2006incentive}: first, customers visit the retailer's website, select their desired stock keeping units (SKUs), and choose a preferred delivery time window during the order acceptance phase. Second, in the order assembly phase, the retailer prepares all accepted orders for the delivery of designated SKUs to customers within the specified time window during the third phase, the order delivery phase. Planning these steps is highly complex, primarily due to the sequential arrival of customers during order acceptance and the inherent demand uncertainty faced by the retailer.

Our paper aims to address the challenges of demand uncertainty in inventory planning for online grocery retailers. Given the highly perishable nature of grocery SKUs, online retailers must make informed inventory decisions in a highly dynamic and stochastic environment to ensure availability upon customer request \citep{winkelmann2022dynamic}. The intense competition in the market further emphasises the importance of satisfying customers, as dissatisfied customers may easily switch to alternative providers once a desired SKU is not available during the booking process. Consequently, retailers often opt for overstocking to mitigate the risk of stock-outs \citep{ulrich2021distributional}. However, overstocking perishable products, such as fruits, vegetables, and meat, can result in significant spoilage with corresponding financial costs for the retailer and environmental consequences.

In this paper, we introduce the idea of subscription offers for e-grocery to mitigate inventory challenges. While subscription-based models are already established successfully in some areas, they are rarely implemented in grocery delivery yet. The basic idea is as follows: customers arrive during order acceptance and fill their shopping basket. Based on the bought SKUs, the retailer makes an offer to deliver some or all of the SKUs on a regular base. The retailer then rewards this commitment by offering a price reduction for the selected SKUs. A similar practice is followed by Amazon through its \textit{Subscribe \& Save} program, where customers can subscribe to convenience products such as cleaning supplies, hygiene products, or pet food. Implementing such a subscription-based model allows the retailer to better anticipate order quantities prior to order acceptance, facilitating more accurate inventory planning. In particular, a share of customer demand becomes deterministic; thus reducing uncertainty by providing advanced demand information \citep{siawsolit2021offsetting}. Incorporating such information is especially beneficial in case of huge uncertainty while addressing very high service level targets as given in e-grocery retailing.

While there are a few publications that touch on the topic of subscriptions in the e-grocery domain, the existing literature (see \citealp{belavina2017online, wagner2021value}) primarily focuses on uniform subscription offers that enhance customer service. However, as uniform subscription offers often harm the retailer's profitability with the narrow profit margins, these papers come to the conclusion that it is challenging to implement profitable subscription offers. However, in these studies, the interplay between implementing product subscriptions and cost savings from improved inventory planning is not considered. In contrast, we propose a three-step procedure that initially enables the retailer to comprehend the costs of uncertainty for each SKU based on buying probabilities and the size of the customer base. Next, we demonstrate how the retailer can quantify the value of gathering additional planning information, such as collecting advanced information on customer demand. Finally, the insights gained from the previous steps can be leveraged to implement subscription offers exclusively for SKUs that enhance the retailer's profitability. This approach ensures the development of profitability-enhancing subscription offers.

\section{Related Literature}

This section provides an overview of related literature. We start by briefly summarising the main characteristics of e-grocery retailing followed by insights into challenges in inventory management within this realm. Finally, we present literature on the benefit of ADI for reducing uncertainty before discussing subscription-based business models presented in the literature and implemented in practice.

\subsection{Characteristics of E-Grocery Retailing}
\label{sec:e-grocery}
e-grocery retailing allows customers to order groceries online to be delivered at a date and time slot determined by the customer. From an operational perspective, it is characterised by the additional fulfilment processes of picking and delivery. Comprehensive reviews of the literature considering logistic strategies in e-grocery retailing are provided by \citep{rodriguez2022grocery, calzavara2023cost}. \citet{winkelmann2022dynamic} present an overview of challenges and opportunities of e-grocery retailing. In particular, while the availability of uncensored demand data and ADI provides opportunities due to new types of data, challenges within inventory management arise from the increased lead time in a dynamic and stochastic environment and the necessity to estimate an extreme (right) quantile of the customer demand distribution to meet high service level targets \citep{ulrich2021distributional}. Additionally, the literature addresses the tasks of warehousing (see e.g.\ \citealp{rodriguez2022grocery}) and attended home delivery (see, e.g.\ \citealp{kohler2023data}).

\subsection{Inventory Management for Grocery Delivery}
\label{sec:inventory}
Inventory management in e-grocery retailing aims to determine replenishment order quantities such that the (expected) sum of costs for inventory holding and spoilage (in case of excess inventory) and shortage costs (in case of excess demand) is minimised. Addressing this trade-off for perishable SKUs is notoriously a huge challenge for retailers that has been discussed in the literature for a long time (see \citealp{nahmias1982perishable} for a review on early literature dealing with ordering policies for perishable inventories). A key difficulty is uncertain customer demand when replenishment order quantities and resulting inventory levels have to be determined. While for some SKUs excess inventory evokes costs for inventory holding only (as remaining units can be sold in the following demand period), grocery retailers offer a variety of fresh SKUs having a very limited shelf life. \citet{karaesmen2011managing} provide an overview of various models addressing the management of perishable and ageing inventories. These models particularly vary in the type of replenishment policy and the consequences of excess demand.

In general, there are two different approaches for the case of excess demand: first, unfulfilled demand can assumed to be lost, and, second, the case of backordering, i.e.\ unfulfilled demand can be accomplished in a subsequent period. Many inventory models rely on the assumption of backordering. In case of periodic review, an order-up-to policy allows to solve this problem to optimality \citep{scarf1960optimality}. \citet{bijvank2011lost} provide a review of inventory systems and state that in practice unfulfilled demand has to be considered as lost in most cases. Thus, for an accurate representation of the environment of retailing, inventory models with lost sales are needed. However, those models require different replenishment policies and are more difficult to solve. At the same time, \citet{huh2009asymptotic} prove asymptotic optimality of order-up-to policies for single-product single-location inventory systems under periodic review. In particular, their approach works well in case of very high service level targets. In another publication, \citet{minner2010periodic} develop a model for inventory control with periodic review. The authors prove that their approach is superior to order-up-to policies under the assumptions of a deterministic shelf life. \citet{winkelmann2022dynamic} provide an extension to previous literature by proposing an approximated dynamic programming approach that allows the incorporation of probability distributions for various sources of uncertainty such as customer demand, shelf life, and potential supply shortages. The authors demonstrate that for an e-grocery retailing business case the impact of stochastic customer demand on resulting costs is highest.

Following this result, for perishable grocery SKUs with very limited shelf life such as fruits and vegetables or meat, the assumptions of the newsvendor model, the classical model in the case of stochastic customer demand with lost sales, are appropriate and allow for a more simple modelling approach \citep{silver1998inventory, zipkin2000foundations}. The newsvendor model assumes a single demand period with a single order decision before customer demand realises. The cost-optimal replenishment order quantity can be obtained by applying a service level $\alpha$ to the (estimated) inverse cumulative distribution function of customer demand. In the newsvendor model, this service level equals the ratio between costs for excess demand per unit and the sum of costs for excess inventory and excess demand per unit. For simplicity reasons, we follow the considerations on demand forecasting for e-grocery retailing practice by \citet{ulrich2021distributional} and rely our analyses in this paper on the assumptions of the newsvendor model. While \citet{qin2011newsvendor} suggest various extensions, this approach is further justified by very recent data-driven approaches on inventory management also relying on the assumptions of the newsvendor model (see, e.g.\ \citealp{lee2021data, xu2022robust}).

\subsection{The Value of Advanced Demand Information}
\label{sec:adi}
Advanced demand information (ADI) refers to customer orders that are placed before the retailer has to determine replenishment order quantities. Thus, a portion of customer demand becomes deterministic instead of stochastic. The period between placing the order and expecting a delivery to arrive is denoted as \textit{demand lead time}. Research in this field dates back to \citep{milgrom1988communication} who demonstrate the substitutional relation between inventory and demand information. \citet{hariharan1995customer} show that the improvement enabled by ADI is in the same manner as a reduction in supply lead times. Since then, various publications have considered the effect of ADI on the profit of a firm in different settings. \citet{gallego2001integrating} mention two different motivations of customers to provide ADI: first, risk-averse customers place their orders early to ensure a higher probability of receiving the desired SKUs on time as some companies give priority to such orders. Second, following \citet{chen2001market}, price-sensitive customers may allow for a larger delay in the delivery when a discount is offered. In our model, we assume that the demand lead time is sufficient to consider subscription-based demand as deterministic.

Another research stream addresses the additional benefit of flexibility in the delivery period. While \citet{karaesmen2004value} demonstrate a significant decrease in costs for manufacturers if customers allow for deliveries before the desired period, \citet{wang2008inventory} obtain the same result within an inventory model when allowing for flexible delivery time windows. In addition to substantial cost reductions, the authors prove that, under these conditions, an increase in the demand lead time is more beneficial than a decrease in the supply lead time. However, in the business model considered in our paper, we assume that the customer can determine a particular day and time slot for delivery. Thus, there is no flexibility in delivery time windows. 

The publication most closely connected to our problem is \citep{siawsolit2021offsetting}. The authors summarise previous literature on ADI and underscore the benefit of ADI for firms by reducing inventory and spoilage and, thus, enhancing profits. At the same time, a portion of these profits can be forwarded to customers who also benefit from ADI through reduced prices. Using a Markov decision process the authors extend previous considerations on ADI by taking into account stochastic customer demand and random shelf lives at the same time. They prove that in case of a demand lead time exceeding the supply lead time, an average increase in profit of 8\% can be reached along with fewer outdated SKUs. This result holds particularly for SKUs with very short shelf lives. The authors finally confirm the applicability of the newsvendor model in the case of a single-period shelf life. In the problem considered in our paper, we rely on the results of \citet{siawsolit2021offsetting} and implement a subscription-based service to gain ADI with the aim of increasing profitability for companies. In contrast to their model, we do not only consider a fixed price discount to attract customers for subscriptions but derive optimal discounts for each customer and SKU individually.

\section{Practical Example of Customer Buying Patterns}
\label{sec:example}

\begin{figure}[t]
\includegraphics[width=1\textwidth]{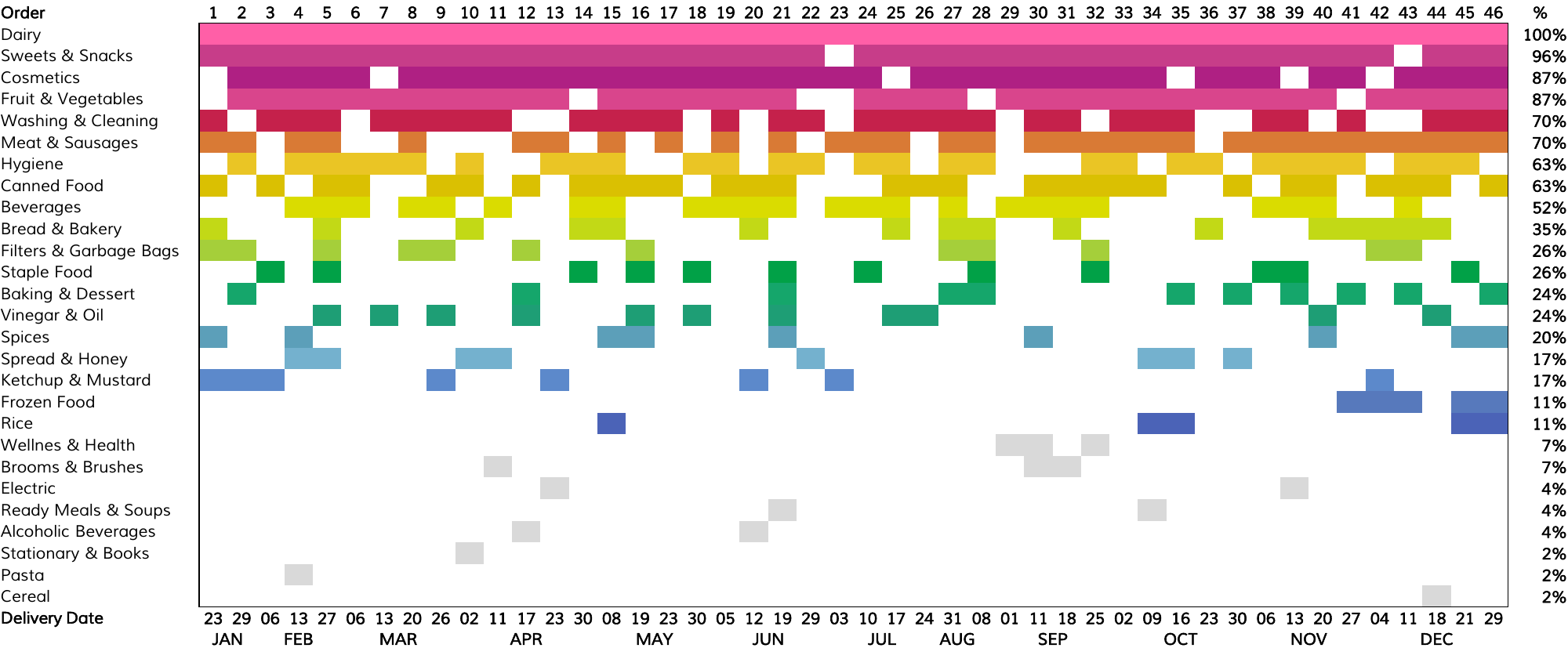}
\caption{Illustrative Example of Ordering Behaviour of a Customer}
\label{fig:customer_orders}
\end{figure}

Before delving into our modelling approach for subscription offers, we present a real-world example to illustrate and contextualize our idea, helping us establish realistic assumptions. Figure~\ref{fig:customer_orders} illustrates a real-world example of customer order behaviour over the course of a year, based on data obtained from an online supermarket in Germany. The figure illustrates the orders and purchased product categories made by a specific customer throughout 2015.

On the top of the x-axis, we represent the order number of that customer for that year, with a total of 46 orders. This indicates that the customer was fairly consistent in placing orders. On the lower part of the x-axis, we show the delivery dates. The customer maintained an almost perfect regular frequency, ordering every week. For instance, in March, the customer placed orders on the 6th, 13th, 20th, and 26th, indicating a new order every 6-7 days. However, there are also deviations from this pattern, such as in August, where only one order was placed on the 8th, likely due to vacation time during that period. The left side of the y-axis represents the product categories of the SKUs included in each order. We colour the corresponding row and cell if an order contains at least one SKU from that product category. For example, in the very first order of that customer, SKUs from nine product categories were ordered, including dairy, sweets \& snacks, washing \& cleaning, meat \& sausages, canned food, bread \& bakery, filters \& garbage bags, spices, and ketchup \& mustard. On the right side of the y-axis, we additionally depict the frequency of these product categories occurring in all the customer's orders.

The figure offers valuable insights into the potential implementation of subscription offers for certain SKUs from the customer's perspective. Firstly, it reveals that certain product categories are consistently purchased, while others are seldom bought. For the former, customers are likely to be highly receptive to subscription offers, particularly if accompanied by price reductions. Conversely, for product categories with infrequent purchases, customers are unlikely to agree to subscriptions or regular deliveries. Secondly, there exist product categories with purchase frequencies falling below 70\% but above 10\%, where ordering patterns are unclear. In such cases, it is uncertain whether customers require SKUs within these categories occasionally, making less frequent subscriptions appealing. Alternatively, customers may purchase these SKUs elsewhere intermittently, or they may consider subscribing if offered an attractive price reduction. In these cases, it is less clear whether offering a subscription would yield success or not.

From the perspective of a retailer responsible for maintaining adequate inventory stock of various SKUs, several conclusions can be drawn. Firstly, estimating the required stock for certain SKUs is more straightforward than for others. This is particularly evident for top-selling SKUs that are consistently purchased in every order. However, as the purchase frequencies decrease and no clear ordering pattern emerges from a customer's historical data, the task of estimating the necessary stock becomes significantly more challenging. Secondly, considering the product categories, there are SKUs for which inaccurate estimation carries a higher cost impact than others. For instance, perishable SKUs like dairy, fruits \& vegetables, and meat \& sausages have relatively shorter shelf lives. Inaccurate estimation for SKUs within these categories could lead to potential waste and increased costs. Conversely, products such as sweets \& snacks, cosmetics, and washing \& cleaning products likely have longer shelf lives, making the cost of misestimation relatively lower than for the first group. Therefore, from a retailer's perspective, it is likely more desirable to offer subscriptions for SKUs where they cannot easily anticipate ordering behaviour and where SKUs are prone to spoilage.

In the following sections, we will leverage these insights to our problem definition and modelling approach. We will integrate customer purchase probabilities for specific SKUs and advocate for the calculation of subscription offers for each product category. This strategy enables retailers to refine their estimates of the costs associated with overstocking certain SKUs. With our approach, retailers can determine, for each purchase probability and associated costs, whether subscription offers should be extended, and if so, the appropriate level of price reductions to maximise overall profitability.

\section{Problem Definition}

We examine the business case of a retailer offering groceries for online purchase. Given the competitive landscape of online groceries, ensuring satisfactory customer service is paramount for retailers. Consequently, they face the challenge of setting the appropriate inventory levels to meet a high service level target.

In the booking process, customers select their desired SKUs on the retailer's website, gradually revealing overall demand. Despite some orders being placed multiple days in advance, the retailer must decide on SKU quantities with respect to a given supply lead time. This poses the challenge of accurately forecasting customer demand when determining replenishment order quantities. In our scenario, the retailer offers subscriptions to customers. These subscriptions come with price discounts if customers agree to order specific SKUs regularly. The retailer makes these offers to gain better knowledge of future demand for certain SKUs. If customers like the offer, they will get SKUs regularly at a reduced price. If not, they can proceed without subscribing. Customer decisions depend on the discount level and the product type, as some SKUs are more suitable for recurring orders than others.

However, due to the narrow profit margins, subscription incentives can only be extended when they are expected to positively impact revenue. Thus, we propose a three-step approach for retailers: in the first step, our objective is to assess the costs of planning uncertainty and their impact on the retailer's profit for specific SKUs. To achieve this, the retailer must analyse historical data to gain insights into customers' purchasing probabilities for various SKUs and their influence on inventory planning challenges. This enables the computation of uncertainty costs for each SKU. In the second step, the retailer focuses on SKUs with significant uncertainty costs and aims to evaluate the potential reduction in these costs by acquiring advanced demand information. This involves determining the extent to which gathering information about intended purchase behaviour from customers could decrease uncertainty costs. For certain SKUs, obtaining information from a limited number of customers may suffice, while for others, persuading a larger proportion of customers to share their purchasing intentions is necessary for a substantial reduction in uncertainty costs to occur. In the final step, all information gathered about the impact of uncertainty costs and the value of advanced demand information can then be translated into profit-increasing subscription offers. 

We always consider customer purchase probabilities when crafting subscription offers. Additionally, we acknowledge that not all customers will opt to subscribe, either due to dissatisfaction with the offer or a general reluctance to commit to a retailer. Our analysis primarily focuses on the retailer's overall profit, particularly in the context of short-term inventory planning solutions. However, we anticipate that successfully encouraging customers to subscribe will lead to enduring positive impacts on the retailer's profitability in the long run.

\newpage
\section{Modelling Approach}

In this section, we outline our modelling approach for assessing the potential of subscriptions to enhance retailer revenue. This process is divided into three steps, each covered by one section: (1) understanding the impact of demand uncertainty and how retailers can comprehend the associated costs, (2) assessing the value of advanced demand information to reduce demand uncertainty, and (3) efficiently leveraging previous results to implement profitable subscription offers. This approach ensures that it can be adapted by any retailer. 

\subsection{The Impact of Demand Uncertainty} \label{section:uncertainty}

We start by detailing how we model uncertainty in customer demand. Afterwards, we derive a function for the (expected) short-term profit of the retailer and discuss the impact of uncertainty in the inventory process on this profit.

\subsubsection{Modelling Demand Uncertainty and Short-Term Profit}
\label{sec:customer_demand}
In our model, we consider a segmentation of potential customers into groups based on their buying probability $\pi$ for the SKU under consideration. For instance, a customer from a single household may purchase cleaning supplies less frequently than a customer from a family household. Hence, all customers from one group have an equal probability of making a purchase, and their buying decisions are assumed to be independent of other customers from that group. The size of the customer base is denoted by $n$, each deciding whether to purchase a single unit of the SKU or not, resulting in a total demand of $X$ (a random variable) for a specific group. For each booking process (each period), the retailer is tasked with determining a replenishment order quantity $q$ for the SKU. Given that each customer independently decides whether to make a purchase or not, the total demand $X$ can be modelled using a binomial distribution with parameters $n$ and $\pi$ \citep{dolgui2008performance}. In cases where $n$ is sufficiently large, this distribution can be approximated by a normal distribution with a mean of $\mu = n \cdot \pi$ and a standard deviation of $\sigma = \sqrt{n \cdot \pi \cdot (1-\pi)}$. Estimating these parameters in retail practice, e.g.\ based on historical data or features, allows forecasting the cumulative distribution function of demand $\hat{F}_X$ \citep{ulrich2021distributional}.

To achieve a formulation for the (short-term) profit of the retailer, we rely on the well-established newsvendor model, a classical framework for modelling stochastic customer demand (see, e.g.\ \citealp{silver1998inventory,zipkin2000foundations}). The newsvendor model operates under the assumption of independent demand periods, where a period typically represents a fixed time interval, such as a day or a week, during which the SKU remains available for sale. Importantly, the model requires the retailer to make a single order decision for a specific quantity of the SKU before the actual customer demand becomes known. Once the order is placed, the retailer cannot make further adjustments to the quantity in this period. For the short-term profit, we do not include strategic long-term considerations but only a revenue $p$ per unit, i.e.\ the additional value of selling the SKU instead of returning or depreciating it after the demand period, and supply costs of $c$ per unit. We assume these values to be fixed in the short run, and thus they do not affect the buying probability of each customer segment. Assuming that remaining units fully depreciate after the period, the ex-post profit $z$ depends on whether the retailer is faced with excess inventory or excess demand in a certain period in the following way:
\begin{equation}
    z= 
    \begin{cases}
        (p-c) \cdot q & \text{if }x>q\\
        p \cdot x - c \cdot q & \text{if }x\leq q.\\
    \end{cases}
    \label{eq:cases}
\end{equation}

As online groceries are in a highly competitive environment, it is especially important for the retailer to provide satisfactory customer service. This results in very high strategic service level targets in e-grocery retailing due to the high and costly risk of long-term customer churn in case of stock-outs. While we are interested in the effect of subscriptions on the short-term profit, we assume the replenishment order quantity $q$ to be chosen such that it satisfies a service level $\alpha$ that is given according to strategic long-term considerations (cf. \citealp{ulrich2021distributional}). In the newsvendor model, this quantity can be observed at the $\alpha$-quantile of the inverse (estimated) cumulative distribution function (CDF) of customer demand: $q = \hat{F}_X^{-1}(\alpha)$. Following the assumption of normally distributed demand and assuming parameters $\mu = n \cdot \pi$ and $\sigma = \sqrt{n \cdot \pi \cdot (1-\pi)}$ to be known enables to apply the corresponding density function $f_{N(\mu, \sigma)}(x)$. The expected profit can then be stated as follows \citep{qin2011newsvendor}:
\begin{equation*}
    E(Z) = p \cdot \int\limits_0^q f_{N(\mu,\sigma)}(x) x \,\, dx + p \cdot q \cdot \left(1-F_{N(\mu, \sigma)} \left(q\right)\right) - c \cdot q.
\end{equation*}
Again, we obtain the two cases given in Equation~(\ref{eq:cases}): the first part represents the revenue if the order quantity exceeds demand (upper case in Equation~(\ref{eq:cases})), and the second represents the case of excess demand (lower case in Equation~(\ref{eq:cases})), both weighted with the corresponding probabilities. Note that the integral can be interpreted as the mean of a normal distributed random variable truncated at $q$ multiplied by the probability that demand does not exceed this quantity, which equals the intended service level $\alpha$. The expected value of this truncated random variable is given by $\mu - \sigma \cdot \frac{f(F^{-1}(\alpha))}{\alpha}$ where $F$ denotes the CDF of the standard normal distribution and $f$ the corresponding probability distribution function. According to the newsvendor model, we obtain the order quantity as $q = \mu + \sigma \cdot F^{-1}(\alpha)$ for a given strategic service level $\alpha$. This translates the expected profit into:
\begin{equation*}
    E(Z) = (p-c) \cdot \mu - p \cdot (f(F^{-1}(\alpha)) - (1-\alpha-c/p) \cdot F^{-1} (\alpha)) \cdot \sigma
\end{equation*}
Substituting $\gamma = p \cdot f(F^{-1}(\alpha)) - (1-\alpha-c/p) \cdot F^{-1} (\alpha)$ and explicitly stating the standard deviation of customer demand allows to rewrite this equation as: 
\begin{equation}
    E(Z) = \underbrace{(p-c) \cdot \mu}_{\text{I: profit without uncertainty (PWU)}} - \underbrace{\gamma \cdot \sqrt{n \cdot \pi \cdot (1-\pi)}}_{\text{II: expected costs of uncertainty (ecu)}}
    \label{eq:profit_base}
\end{equation}

This formulation enables separating the expected profit into two parts: Part I corresponds to the \textit{profit without uncertainty (PWU)} in case of perfect information ($\sigma = 0$), i.e.\ if demand would realise at its expected value and the retailer would know this before the replenishment order quantity needs to be determined. PWU increases linearly in the number of customers $n$, the individual buying probability $\pi$, and the short-term margin $(p-c)$. Part II represents \textit{expected costs of uncertainty (ecu)} as units can not be transferred to the following period and negatively affect the expected profit.

\subsubsection{Exploring the Factors Influencing Expected Costs of Uncertainty}
\label{sec:impact_uncertainty}

The profit function derived in Equation~(\ref{eq:profit_base}) reveals that the expected profit is negatively affected by the presence of uncertainty (induced by the variance parameter $\sigma$). Therefore, we are particularly interested in understanding how the customer base and their respective buying probabilities for an SKU impact these uncertainty-related costs. It's worth noting that we encounter retailers of varying scales in terms of business reach, with newer online retailers potentially having smaller customer bases. Additionally, different products may exhibit varying buying probabilities. Consequently, we aim to explore the individual effects of these factors on overall profit. The retailer would be particularly interested in reducing uncertainty by subscription in cases where those costs diminish the expected profit comprehensively.

In Table~\ref{tab:uncertainty}, we present calculations for various scenarios. For the sake of simplicity, we assume a unit selling price of $p=1$, supply costs of $c=0.85$, and an intended service level of 97\% (which is in line with the assumptions of \citealp{ulrich2021distributional}). For $n$, the settings are in accordance with the research findings of \citep{kohler2023data}, who report an average of approximately 60 expected daily orders for a typical German online retailer. Note that in our case one demand period could be, e.g.\ one week. To provide insights that cover a broader customer base, we present results for different values of $n$, ranging from 50 to 1,000 customers. It's important to clarify that our use of the variable $n$ specifically refers to the number of potential customers for a particular SKU within a given customer group rather than representing the entire customer base of the retailer. We also explore a range of values for $\pi$, which varies from 1 (indicating constantly purchasing the SKU) to 0.25 (suggesting SKUs are purchased with only a 25\% probability of occurrence).

\begin{table}[t!]
    \centering
    \scalebox{0.8}{
    \begin{tabular}{c|rr|rr|rr|rr}
         & \multicolumn{2}{c|}{$\pi$ = 1.0} & \multicolumn{2}{c|}{$\pi$ = 0.75} & \multicolumn{2}{c|}{$\pi$ = 0.5}  & \multicolumn{2}{c}{$\pi$ = 0.25}\\
         \hline
         & PWU & ecu & PWU & ecu & PWU & ecu & PWU & ecu \\
        \hline
        $n$ = 50 & 7.50 & 0 & 5.63 & 4.93  & 3.75 & 5.69  & 1.88 & 4.93 \\
        $n$ = 100 & 15.00 & 0  & 11.25 & 6.97  & 7.50 & 8.05  & 3.75 & 6.97\\
        $n$ = 500 & 75.00 & 0   & 56.25 & 15.59  & 37.50 & 18.00  & 18.75 & 15.59\\
        $n$ = 1000 & 150.00 & 0  & 112.50 & 22.05  & 75.00 & 25.46  & 37.50 & 22.05 \\
    \end{tabular}}
        \caption{Profit without uncertainty (PWU) and expected costs of uncertainty (ecu) for different values of the customer base $n$ and the buying probability $\pi$ for given selling price $p = 1$, supply costs $c = 0.85$ and service level $\alpha = 0.97$.}
    \label{tab:uncertainty}
\end{table}

Table~\ref{tab:uncertainty} gives results for both Part I and Part II of Equation~(\ref{eq:profit_base}), totalling the profit deductions due to uncertainty. The first column refers to a buying probability of 1, signifying a scenario with no uncertainty (where every customer always purchases the SKU). This serves as a baseline for our analysis. However, as demonstrated in the earlier example in Section~\ref{sec:example}, we have observed that certain SKUs may also be consistently included in a customer's shopping basket. For this scenario, the profit range is between 7.5 and 150 depending on the number of customers considered. If we now turn towards the scenarios with a lower buying probability we can see that PWU decreases (as customers are less likely to buy). ECU takes the highest values for buying probabilities close to $\pi = 0.5$. In the scenario of $\pi=0.75$, the expected profit is still positive for all settings of $n$, while ECU already make up a large portion of the revenue. For $\pi = 0.5$ the expected profit becomes negative considering 50 and 100 customers. Even when the profit is still positive for larger customer bases, ECU lead to a reduction of almost 50\% with $n=500$ and a third for $n=1.000$. For $\pi = 0.25$ (last column) results are even worse.

\begin{figure}[ht]
    \centering
    \includegraphics[width=0.6\textwidth]{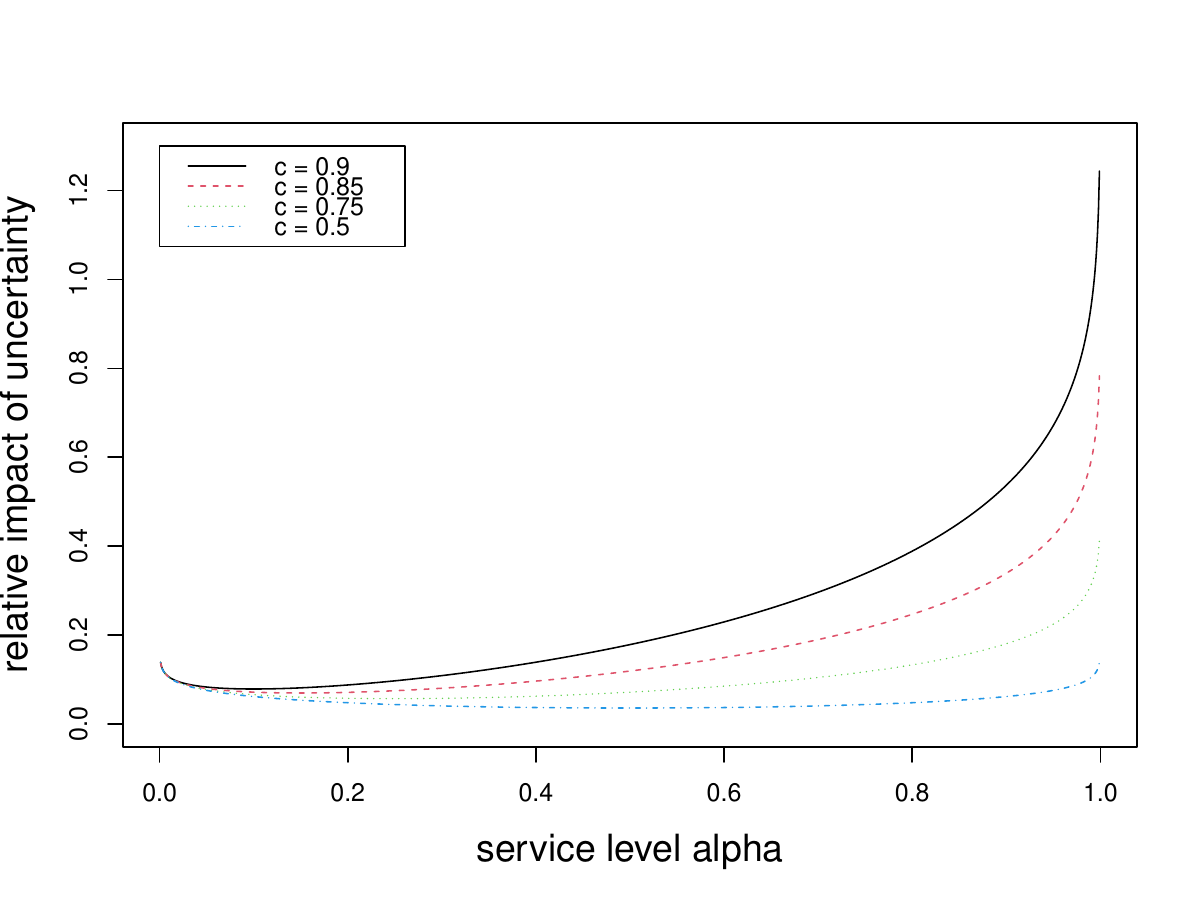}
    \caption{Ratio ecu/PWU between expected costs of uncertainty and the profit without uncertainty depending on supply costs $c$ and service level $\alpha$ for selling price $p = 1$, customer base $n = 500$, and buying probability $\pi = 0.5$.}
    \label{fig:alpha_gamma}
\end{figure}

e-grocery retailing is characterised by small profit margins and high service level targets. Figure~\ref{fig:alpha_gamma} gives the relation between the service level $\alpha\in(0,1)$ and the relative difference between PWU and the expected profit due to ECU on the y-axis for different values for supply costs $c\in\{0.5, 0.75, 0.85, 0.9\}$ indicated by colours. We hold the selling price $p=1$, the customer base $n=500$, and the buying probability $\pi=0.5$ constant. For instance, for a service level of 50\% (i.e.\ without any safety stock) and supply costs $c=0.85$, we obtain that uncertainty leads to a decrease in the expected profit of less than 20\%. The impact of uncertainty increases for very large service level targets and is higher for smaller margins. Note that for $c = 0.9$ and very high service level targets, ECU exceed PWU even for the $n = 500$ scenario, leading to a negative expected profit.

We have delved into modelling expected profit and uncertainty costs in this section. Our analysis is centred around individual SKUs and customer segments, making it readily applicable for retailers to extend this modelling to each of their respective groups. This involves calculating PWU and ECU using their own data related to customer demand, desired service levels, purchase probabilities, prices, and costs. In general, the following insights can be derived:

\begin{itemize}
\item ECUconsiderably impact the overall expected profit. This holds particularly for very high service level targets and small margins as prevalent in e-grocery retailing.
\item In some cases, ECUmay even eat up PWU, resulting in a negative expected profit overall.
\item ECU are especially challenging for smaller customer bases. 
\item Considering the absolute values of ECU, we find that they are highest for the $\pi = 0.5$ scenario (and equal for $\pi = 0.75$ and $\pi = 0.25$). Hence, retailers dealing with uncertainty are particularly challenged when it comes to SKUs with buying probabilities around 50\%.
\end{itemize}

\subsection{The Value of Advanced Demand Information}
\label{sec:advanced_demand}
As we can obtain from Equation~(\ref{eq:profit_base}), a higher variance of customer demand increases ECU and, therefore, leads to a stronger reduction in the expected profit of the retailer. Having advanced demand information (ADI), i.e.\ information about customers' order intentions already before the booking process, can help reduce uncertainty in customer demand and, thus, increase the expected profit of the retailer. Our primary focus in this section is analysing how much ADI can contribute to increased profitability ceteris paribus, i.e.\ without changing the value of other model parameters such as the selling price. This analysis serves as input to understand the necessary level of incentives required to achieve this level of ADI.

Gathering ADI could be as simple as a short survey presented at the end of the booking process, wherein customers are asked if they will order a certain SKU again next week. We assume that a share $\beta$ of customers is willing to commit that information and actually behaves as indicated in the survey. Thus, we can separate the customer base for the next booking process into two groups: (1) a \textit{deterministic} group where we know how many customers will buy the SKU (based on their information given in the survey) and (2) a \textit{stochastic} group of customers that did not respond to the survey or did not buy the SKU in the previous week and where we do not have any ADI. In expectation, group (1) comprises $n \cdot \pi \cdot \beta$ customers. A share $\pi$ of these customers is expected to reply that they will buy the SKU again; a share $1-\pi$ will not buy again. On the other hand, group (2) consists of $n \cdot (1-\pi \beta)$ customers buying with a probability of $\pi$. This leads to the expected profit for the second booking process:
\begin{equation}
    E(Z_2) = \underbrace{(p-c) \cdot \mu}_{\text{I: PWU}} - \underbrace{ \gamma \cdot \sqrt{n \cdot (1-\pi\beta) \cdot \pi \cdot (1-\pi)}}_{\text{II: ecu}}
    \label{eq:benefit}
\end{equation}

\begin{table}[t]
    \centering
    \scalebox{0.8}{
    \begin{tabular}{cc|rrr|c}
        $\pi$ & $\beta$ & $Z_1$ & $Z_2$ & $\Delta Z$& relative increase \\
        \hline
        0.25 & 0.25 & 3.15 & 3.65 & 0.50 & 15.68\% \\
        0.25 & 0.5 & 3.15 &  4.17 & 1.01 & 31.88\% \\
        0.25 & 0.75 & 3.15 & 4.70 & 1.54 & 48.68\% \\
        0.25 & 1 & 3.15 & 5.25 & 2.09 & 66.14\% \\
        \hline
        0.5 & 0.25 & 19.50 & 20.66 & 1.16 & 5.96\% \\
        0.5 & 0.5 & 19.50 & 21.91 & 2.41 & 12.37\% \\
        0.5 & 0.75 & 19.50 & 23.27 & 3.77 & 15.88\% \\
        0.5 & 1 & 19.50 & 24.77 & 5.27 & 27.05\% \\
        \hline
        0.75 & 0.25 & 40.66 & 42.20 & 1.54 & 3.78\% \\
        0.75 & 0.5 & 40.66 & 43.92 & 3.27 & 8.03\% \\
        0.75 & 0.75 & 40.66 & 45.94 & 5.28 & 12.98\% \\
        0.75 & 1 & 40.66 & 48.45 & 7.80 & 19.17\% \\
    \end{tabular}}
        \caption{Increase in the expected profit in the second booking process with advanced demand information compared to the first phase for a customer base of $n = 500$ depending on the buying probability $\pi$ and the share of customers responding $\beta$.}
    \label{tab:profit_information}
\end{table}

Expected customer demand and, therefore, PWU, remain constant for the second phase. However, ECU decrease with a smaller variance, i.e.\ more customers replying to the survey (increase in group 1 and corresponding decrease in group 2). In Table \ref{tab:profit_information}, we utilise the formulated approach to assess the value of having ADI available by comparing the profit for the initial booking process $E(Z_1)$ with that for the subsequent booking process $E(Z_2)$. The table consists of six columns. In the first columns, we vary the settings for the buying probabilities $\pi$ and the proportion of customers willing to provide ADI, denoted as $\beta$. We then present the resulting expected profits and the difference between these profits. The final column gives the relative increase in expected profit between the first booking process without ADI and the second booking process with ADI.

Upon examining the table, we observe two predominant effects: first, as the value of $\beta$ increases, indicating more customers are willing to provide ADI, our expected profits increase as expected. Second, the impact of lower buying probabilities on the difference in expected profits is particularly evident. The uncertainty surrounding whether an SKU will be purchased in the following week significantly amplifies the importance of gathering information. It is also noteworthy that all levels of ADI aid the retailer in profit augmentation. However, the higher the overall proportion of customers willing to provide ADI, the greater the expected profit. Therefore, when contemplating offering price reductions, the retailer can offer higher incentives when there is a chance to persuade nearly all customers. Conversely, when only some customers can be convinced to share information or subscribe to a service, lower incentives should be provided to avoid diminishing the additional profit gained.

Summarising, we find that ADI given by customers informing the retailer on their buying decision for the next period before the replenishment order quantity needs to be determined allows for reducing ECU leading to a higher expected profit. This holds particularly if the buying probability and the associated expected profit are low, with a corresponding strong reduction in the expected profit by ECU, and the share of customers providing information is high.

\subsection{The Implementation of Subscription Offers} 
\label{section:subscriptions}
The previous section outlined the impact of uncertainty on the expected profit of the retailer and demonstrated the potential of ADI to reduce this uncertainty. However, in most practical cases, the retailer has to reward a benefit to customers to receive ADI. We identified subscription offers as a possible tool to gather ADI by providing a certain discount for the commitment. To look at the effect of subscriptions for certain customers and SKUs, we (1) set up the short-term profit function in the case of subscription, (2) discuss the effect of subscription on the expected profit for different scenarios, and (3) derive the optimal discount for customers who agree to the subscription offer.

\subsubsection{The Profit in Case of a Subscription-Based Service}
\label{sec:profit_sub}
In the following, we introduce a subscription-based model that allows the retailer to offer a reduced price to customers who commit to buying the SKU in one to multiple following periods. This commitment to ordering one unit of the SKU takes place before the retailer needs to determine the replenishment order quantity. If the customer accepts the subscription offer, the retailer offers a discount $\tau$, reducing the price of the SKU to $p - \tau$. We denote the portion of customers committing to the subscription-based service, i.e.\ the part of customer demand becoming deterministic, by $\beta$ again, depending on the benefit rewarded, now: $\beta = \beta(\tau)$. Agreeing to the subscription offer increases the buying probability of those customers to $1$, rendering this part of demand deterministic. The demand of remaining customers is still stochastic with probability $\pi$. Thus, expected customer demand increases under subscription offers compared to Equation~(\ref{eq:profit_base}). The corresponding variance is only affected by the number of customers ordering at random and the buying probability. As we still take into account strategic considerations and attempt to fulfil a proportion $\alpha$ of uncertain customer demand, we can still assume the order quantity to be given by the newsvendor model. Specifically, this quantity is chosen such that it addresses the $\alpha$-quantile of the inverse cumulative distribution function of stochastic customer demand again along with the complete deterministic demand. This translates the expected profit with subscription offers into the following equation:
\begin{equation}
    \begin{split}
    E(Z_{\text{sub}}) = & \underbrace{(p-\tau-c) \cdot n \cdot \beta(\tau)}_{I_{det}} + \underbrace{(p-c) \cdot n \cdot (1-\beta(\tau)) \cdot \pi}_{I_{stoch}}\\
     & - \underbrace{\gamma \cdot \sqrt{n (1-\beta (\tau)) \pi \cdot (1-\pi)}}_{II: ecu}
    \end{split}
    \label{eq:profit_sub}
\end{equation}

While under ADI in Equation~(\ref{eq:benefit}) PWU (Part I) is equal to the situation without ADI (see Equation~(\ref{eq:profit_base})), when introducing subscription offers, PWU is composed of two parts, $I_{\text{det}}$ and $I_{\text{stoch}}$ and potentially deviates from the previous value. Whether this deviation is positive or negative depends on the discount offered $\tau$ and the share of customers subscribing $\beta$.
Part II again accounts for ECU and reduces with a smaller variance of customer demand, i.e.\ more customers subscribing, due to ADI provided to the retailer.
The situation of perfect information is now given in case all customers would agree to the subscription offer ($\beta = 1$). The variance is equal to 0 then leading to a total (deterministic) profit given as $(p - \tau - c) \cdot n$. On the other hand, in the situation without any customer subscribing ($\beta = 0$), the expected profit reduces to Equation~(\ref{eq:profit_base}).

\subsubsection{The Impact of Model Parameters on the Benefit of Subscription}
As discussed before, the introduction of subscription affects (1) PWU and (2) ECU. The impact on (2) is positive in every case; however, the effect of subscription offers on (1) depends on the discount $\tau$ and the share of customers agreeing to the subscription offer $\beta$. In particular, this effect is positive if the difference $\delta$ between the discount and the SKU of the short-term margin $(p-c)$ and the probability that a customer ordering at random does not buy the SKU $(1-\pi)$ is negative:
\begin{equation}
    \delta = \tau - (1-\pi) \cdot (p-c) \leq 0
    \label{eq:tau}
\end{equation}

We now provide numerical examples and implications for the practice of e-grocery retailers. As a reference, we refer to our basic scenario considering $n = 500$ potential customers, an individual buying probability of $\pi = 0.5$, normalised price $p = 1$, and supply costs $c = 0.85$. For the first set of analyses, we assume a price discount of $\tau = 7.5\%$ and $\beta = 10\%$ of customers agreeing to the subscription offer. This price discount leads to the situation where PWU does not depend on the share of customers subscribing $\beta$ (as $\delta = 0$) and is equal to PWU without subscription (see part I of Equation~(\ref{eq:profit_base})). Thus, the introduction of subscription offers affects the expected profit only by reducing ECU. In the following, we elaborate on the effect of (1) the base of potential customers $n$, (2) the effect of the individual buying probability $\pi$, (3) the effect of the margin determined by the supply costs $c$, and (4) the subscription-related parameters $\tau$ and $\beta$ on the expected profit; as well as (5) the marginal effect of an additional customer subscribing for this setting.

\subsubsection*{Implication 1: The effect of potential customers $n$}

\begin{figure}[htp!]
    \centering
    \subfigure[customer base vs.\ expected costs of uncertainty (ecu)]{\includegraphics[width=0.42\textwidth]{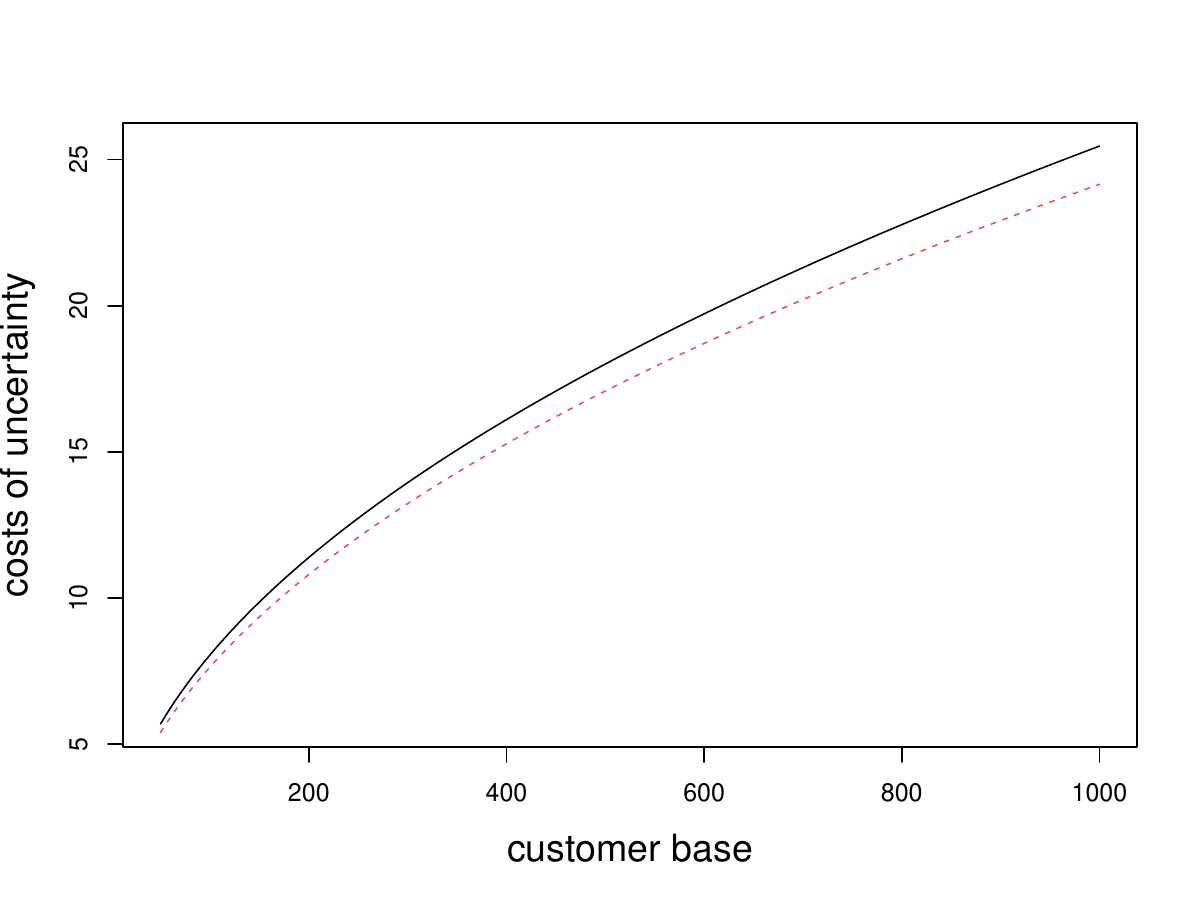}}
    \subfigure[customer base vs.\ relative increase in expected profit]{\includegraphics[width=0.42\textwidth]{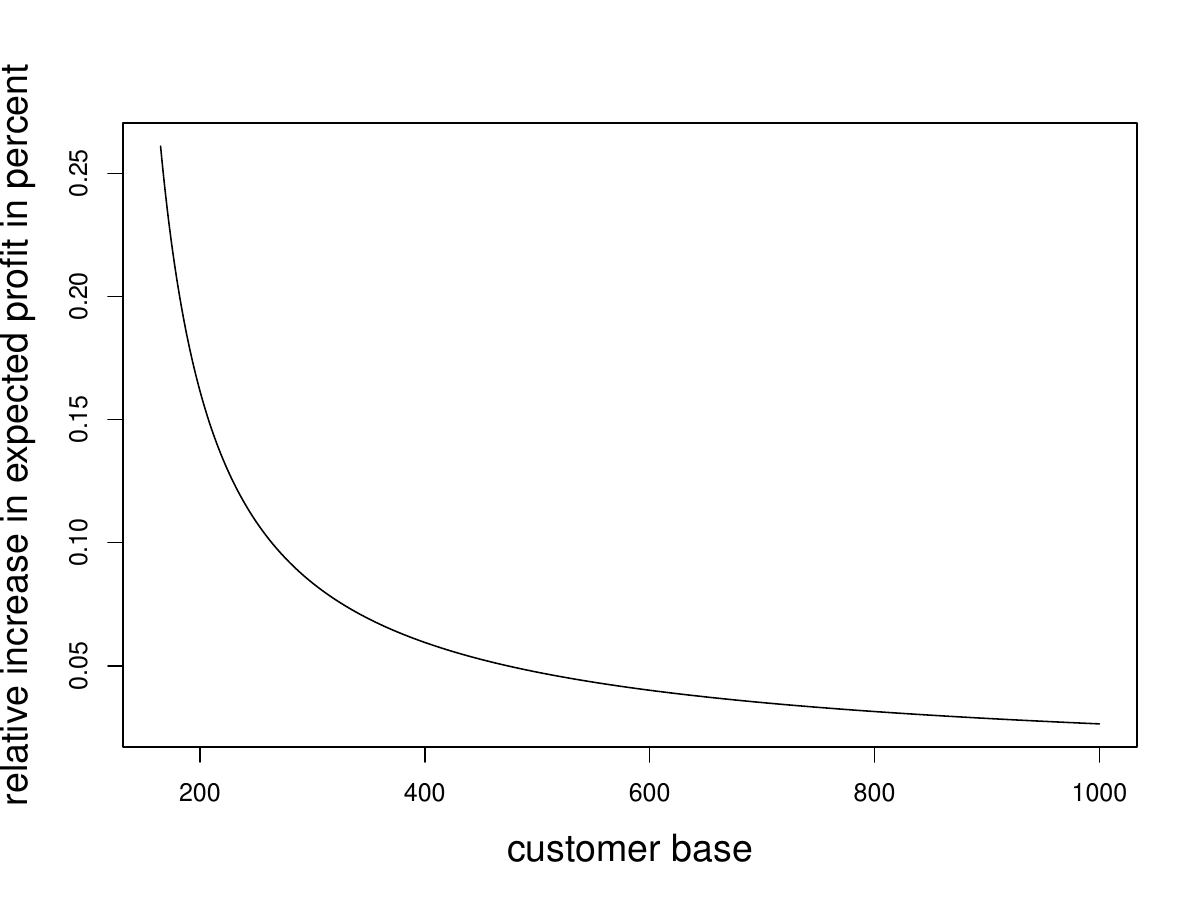}}
    \caption{Relation between the customer base and expected costs of uncertainty (left panel) in the setting without subscription (black line) and with a subscription offer (red dotted line) as well as the relative increase in the expected profit under the subscription-based model (right panel) for buying probability $\pi = 0.5$.}
    \label{fig:effect_n_sub}
\end{figure}

First, we consider how a change in the base of potential customers $n\in[165, 1000]$ impacts both, costs of uncertainty and the relative increase in the expected profit under the subscription model. Figure~\ref{fig:effect_n_sub} gives the relation between the customer base and ECU (left panel) in the setting without subscription (black line) and with subscription offers (red dotted line) as well as the relative increase in the expected profit under the subscription-based model (right panel). Note that we limit the considerations to cases with positive expected profit even without subscription offers and resulting ADI.

As shown above, PWU is equal to the setting without subscription offers. The expected profit is only affected by the decrease in ECU. We find that the absolute decrease in these costs increases with a larger customer base. This is driven by the relation between $n$ and ECU dominated by the square root function and aligns with our findings in Section~\ref{sec:advanced_demand}. However, as the expected profit is smaller for a smaller customer base, the relative increase in the expected profit under the subscription-based model is higher for a smaller customer base. This implies that the subscription offer is particularly beneficial for SKUs with a smaller customer base.

\subsubsection*{Implication 2: The effect of the individual buying probability $\pi$}

\begin{figure}[htp!]
    \centering
    \subfigure[buying probability vs.\ costs of uncertainty]{\includegraphics[width=0.42\textwidth]{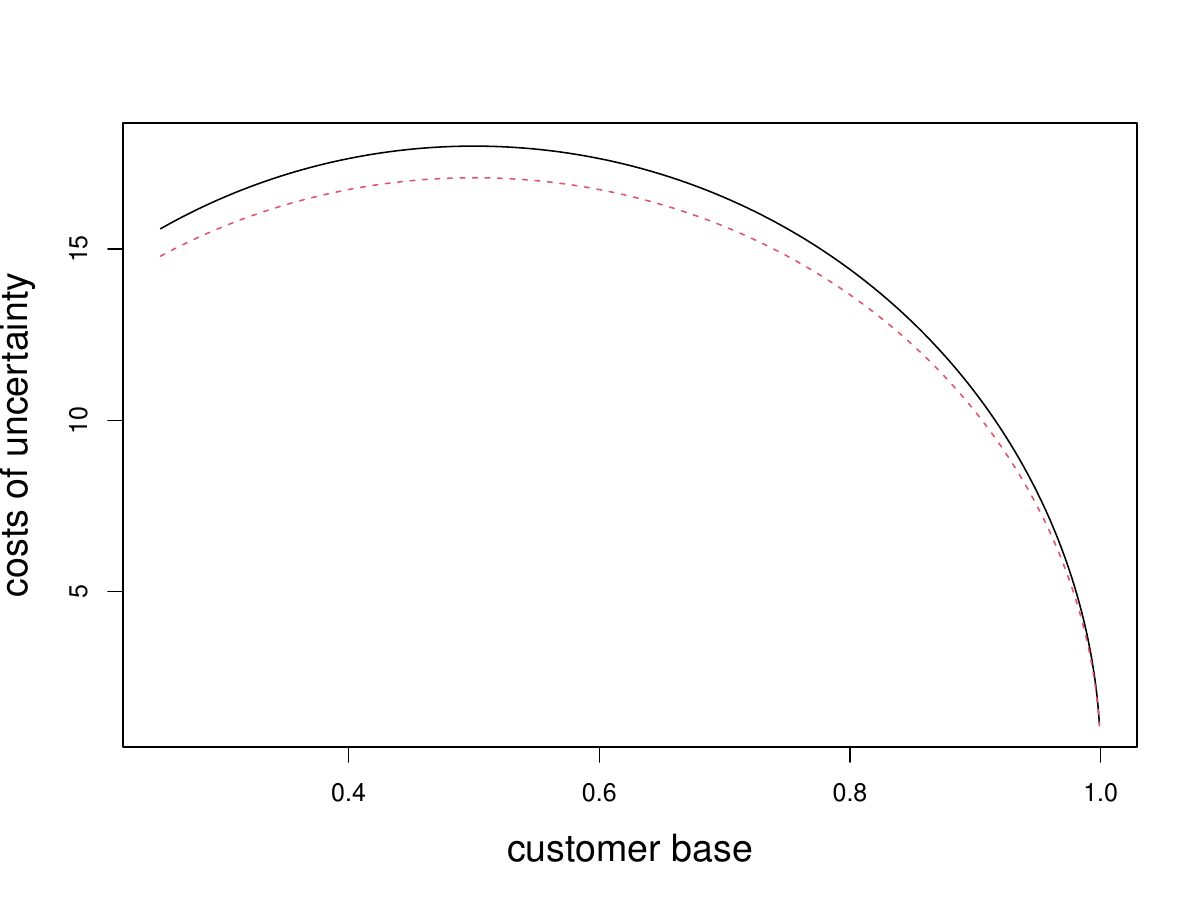}}
    \subfigure[buying probability vs.\ relative change in expected profit]{\includegraphics[width=0.42\textwidth]{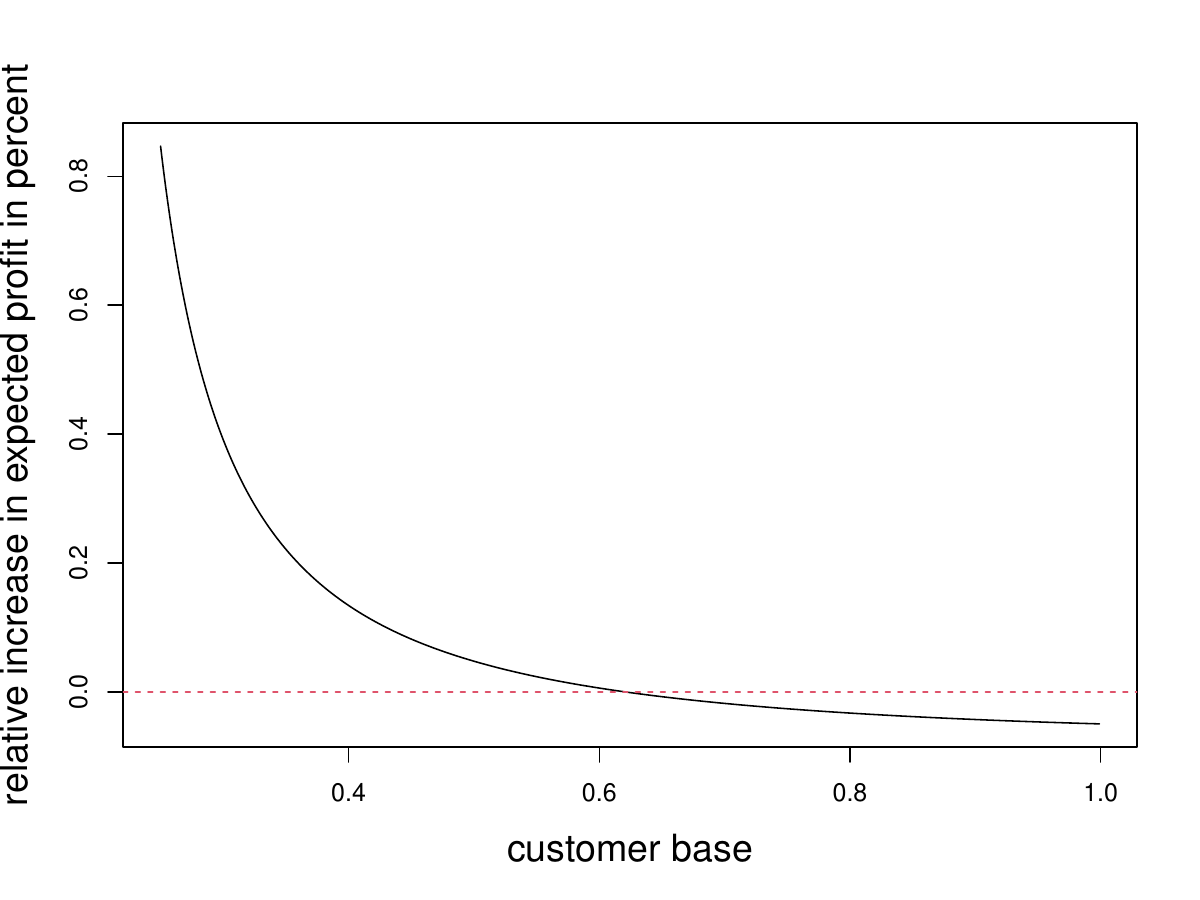}}
    \caption{Relation between buying probability and costs of uncertainty (left panel) in the setting without subscription (black line) and with a subscription offer (red dotted line) as well as the relative increase in the expected profit under the subscription-based model (right panel) for customer base $n = 500$.}
    \label{fig:effect_pi_sub}    
\end{figure}

We now consider the effect of the individual buying probability $\pi$. Figure~\ref{fig:effect_pi_sub} (left panel) illustrates ECU depending on buying probabilities of larger than 0.25 for the baseline scenario (black line) as well as for the subscription-based model (red dotted line). Note that we again consider customer bases with positive expected profit in the situation without subscription offers only. The right panel again gives the relative change in the expected profit depending on the buying probability. In this case, PWU deviates from the situation without subscription offers. Specifically, it is larger in the subscription case for $\pi<0.5$ and larger in the baseline case for $\pi>0.5$.

As a first observation, we find that the retailer won't offer the SKU if the individual buying probability in the situation without subscription offers is fairly small (below 18.8\% for the setting considered in this example). The absolute decrease in ECU is highest for $\pi = 0.5$, i.e.\ if costs are also highest. As PWU increases for higher values of $\pi$, the relative benefit of subscription offers becomes smaller, too. In fact, we find a threshold $\pi^{\text{crit}} = 0.62$ after which a subscription offer with a discount of 7.5\% and 10\% customers accepting this offer is not beneficial for the retailer anymore (indicated by the red dotted line corresponding to the same expected profit with and without subscription offers). This is driven by two aligned effects: (1) the increase in the expected number of customers buying the SKU under the subscription offer is smaller if the buying probability is already high and (2) the benefit regarding ECU also becomes smaller with buying probability approaching 1 (see Figure~\ref{fig:effect_pi_sub}~(a)). Thus, this benefit does not outweigh the loss due to offering the discount $\tau$ for subscribing customers leading to a decrease in the expected profit.

\subsubsection*{Implication 3: The effect of the supply costs $c$}

\begin{figure}[ht]
    \centering
    \includegraphics[width=0.6\textwidth]{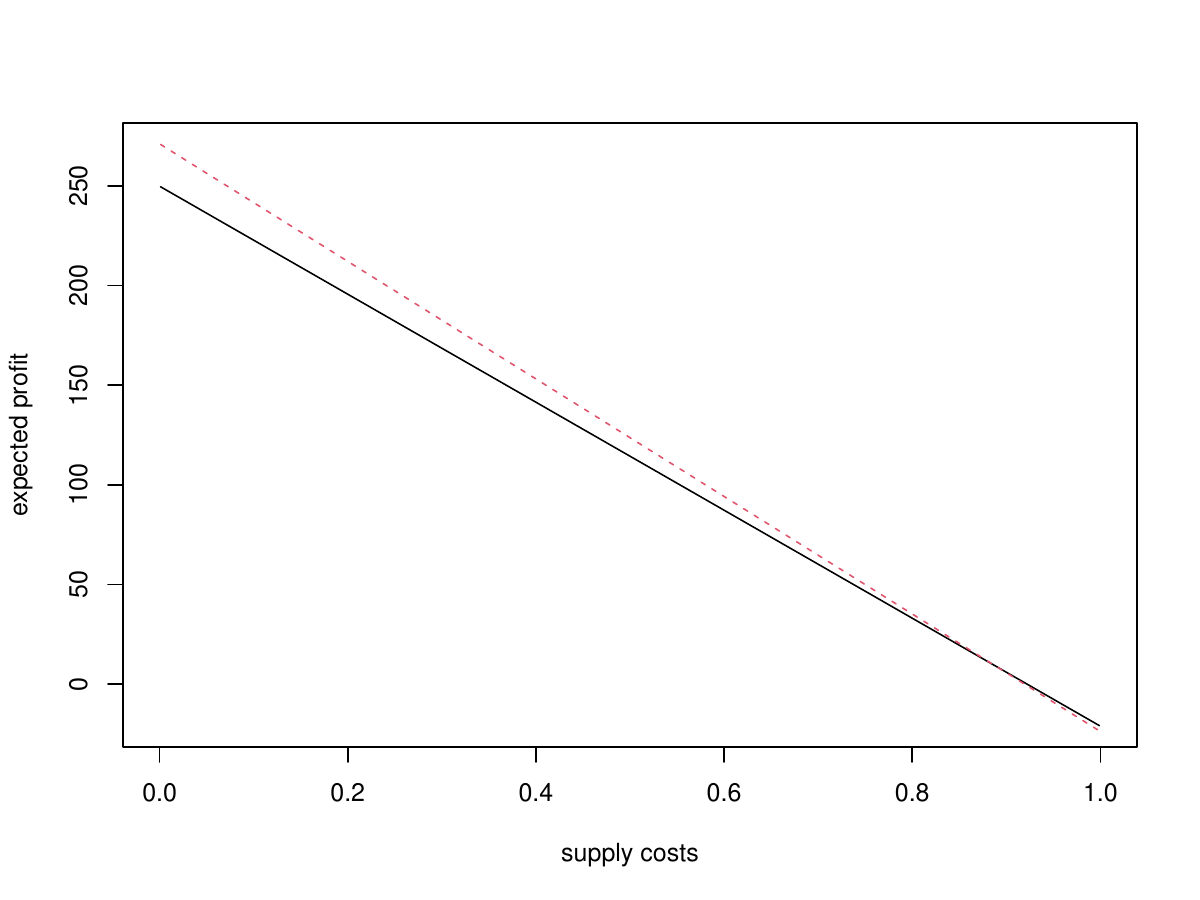}
    \caption{Relation between the supply costs $c$ and the expected profit for the baseline model (black solid line) and the model with subscription offers (red dotted line)}
    \label{fig:c_effect}
\end{figure}

In the third step, we consider the impact of the supply costs $c$ on the expected profit and the impact of uncertainty. As we normalise the selling price to $p = 1$, we consider different values $c\in (0,1)$. Figure~\ref{fig:c_effect} provides the relation between the supply costs and the expected profit for the baseline model (black line) as well as the model with subscription offer (red dotted line). 

Clearly, the expected profit for both the baseline and the subscription case decreases with an increase in supply costs $c$. Specifically, the expected profit becomes negative in the situation without subscription for $c \geq 0.922$. Due to the discounted price for customers accepting the subscription offer, there is a threshold $c^{\text{crit}}$ after which a subscription offer is not beneficial for the retailer anymore, i.e.\ the expected profit without subscription is higher. For the numerical example considered here, we find $c^{\text{crit}} = 0.889$ and a negative expected profit for $c \geq 0.92$ with subscription offers. These values are of course dependent on the other model parameters, in particular the discount $\tau$ and the share of customers subscribing $\beta$.

\subsubsection*{Implication 4: Impact of the customer share committing to the subscription-based service $\beta$}

\begin{figure}[ht]
    \centering
    \includegraphics[width=0.6\textwidth]{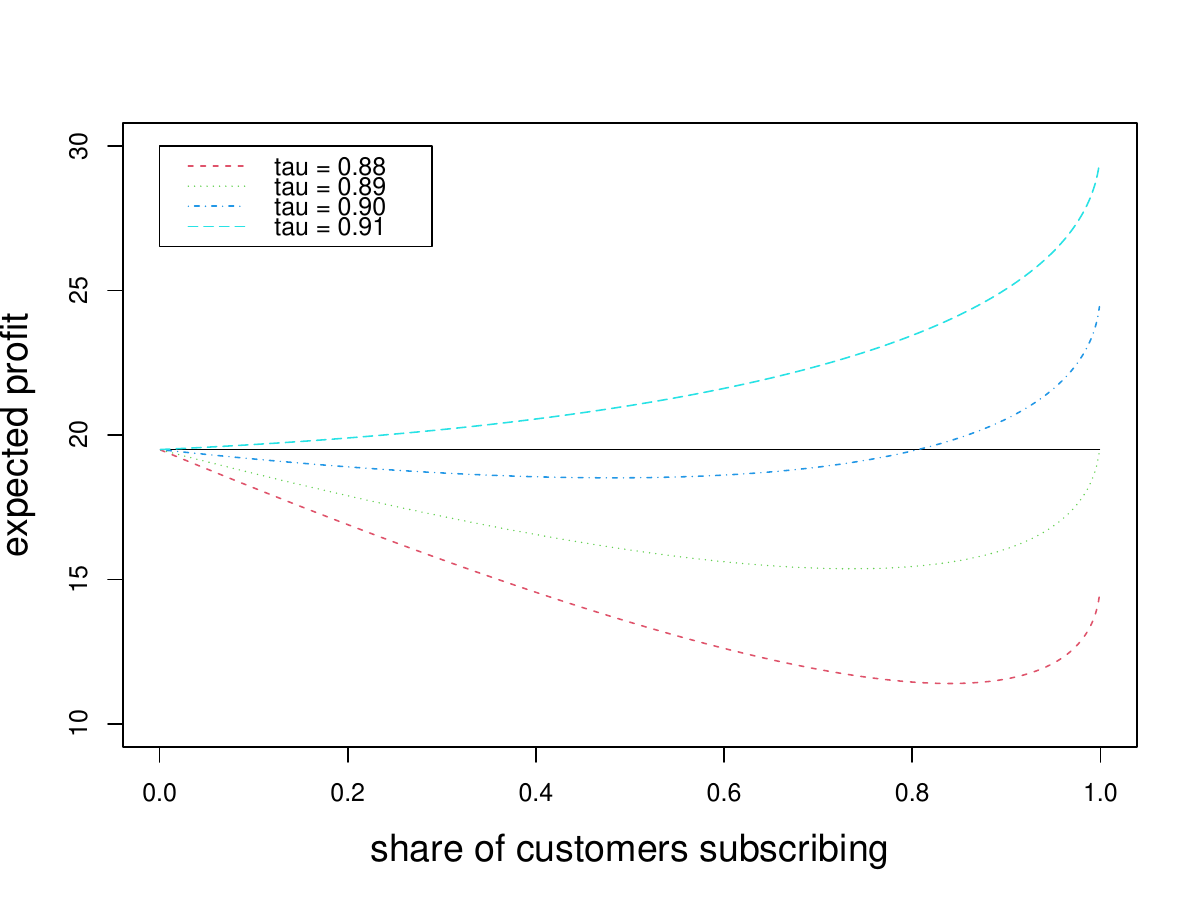}
    \caption{Relation between the share of customers subscribing $\beta$ and the expected profit for different values of the discount $\tau \in \{0.09, 0.10, 0.11, 0.12\}$.}
    \label{fig:beta_sub}
\end{figure}

In the following, we analyse the effect of the share of customers subscribing on the expected profit for different values of the price discount. As discussed above, margins in e-grocery retailing are fairly small and the decision on the discount is a crucial task. Therefore, we vary this discount only slightly within the numerical experiments: $\tau \in \{0.09, 0.10, 0.11, 0.12\}$. The results are illustrated in Figure~\ref{fig:beta_sub}, the black line corresponds to the profit without subscription offers.

We find that the profitability of subscription offers depends on both the discount $\tau$ and the share of customers subscribing $\beta$. For $\tau = 0.12$ and $\tau = 0.11$ the expected profit is lower in the subscription case while it is higher for $\tau \leq 0.09$ regardless of the share of customers subscribing. The case of $\tau = 0.10$ is particularly interesting as offering subscriptions is only beneficial if the retailer can convince a large share of customers to subscribe. While for $\tau = 0.11$ the expected profit closely approaches the expected profit without subscriptions, we find a threshold for the share of customers subscribing of $\beta^{\text{crit}} = 0.807$ to exceed the expected profit without subscriptions in case of $\tau = 0.10$.

\subsubsection*{Implication 5: The marginal effect of an additional customer subscribing}

Motivated by the convex relation between the share of customers subscribing and the expected profit obtained from the results on Implication~4, we analyse the marginal benefit of an additional customer agreeing to the subscription offer. This forms the basis for deriving benefits offered to customers for the commitment. We derive the marginal effect of the expected profit with respect to the share of customers already subscribed $\beta$. As discussed above, introducing a subscription affects the expected profit in multiple ways. As the expected profit function is of an additive form, we can consider the derivatives of each part separately: $\frac{\partial E(Z_{sub})}{\partial \beta} = \frac{\partial \text{I}_{\text{det}}}{\partial \beta} + \frac{\partial \text{I}_{\text{stoch}}}{\partial \beta} + \frac{\partial II}{\partial \beta}.$ The marginal effect of part $\text{I}_{\text{det}}$ gives the increase in the expected profit by an additional customer ordering deterministically at the discount $\tau$. The second marginal effect accounts for the decrease in the expected profit for one customer less ordering randomly with probability $\pi$ at price $p$. Lastly, the increase in the expected profit by reduced ECU is covered by the third marginal effect. The sum of the first two effects is simply given by Equation~(\ref{eq:tau}), multiplied by the customer base $n$; the last effect is given as $\frac{\partial II}{\partial \beta} = \frac{\gamma}{2} \sqrt{\frac{n \cdot \pi \cdot (1-\pi)}{1-\beta}}$. Accumulating these effects gives the marginal effect of an additional customer subscribing:
\begin{equation}
    \frac{\partial E(Z_{\text{sub}})}{\partial \beta} = \delta \cdot n + \frac{\gamma}{2} \sqrt{\frac{n \cdot \pi \cdot (1-\pi)}{1-\beta}}.
    \label{eq:marginal_effect}
\end{equation}

Remember that the first part is positive if the discount is chosen such that the subscription offer is beneficial in every case. In other cases, the marginal effect is only positive for those values of $\beta$ where the reduction in ECU in case an additional customer subscribing exceeds the reduction in part I of the expected profit function.

\begin{figure}[htp!]
    \centering
    \includegraphics[width=0.6\textwidth]{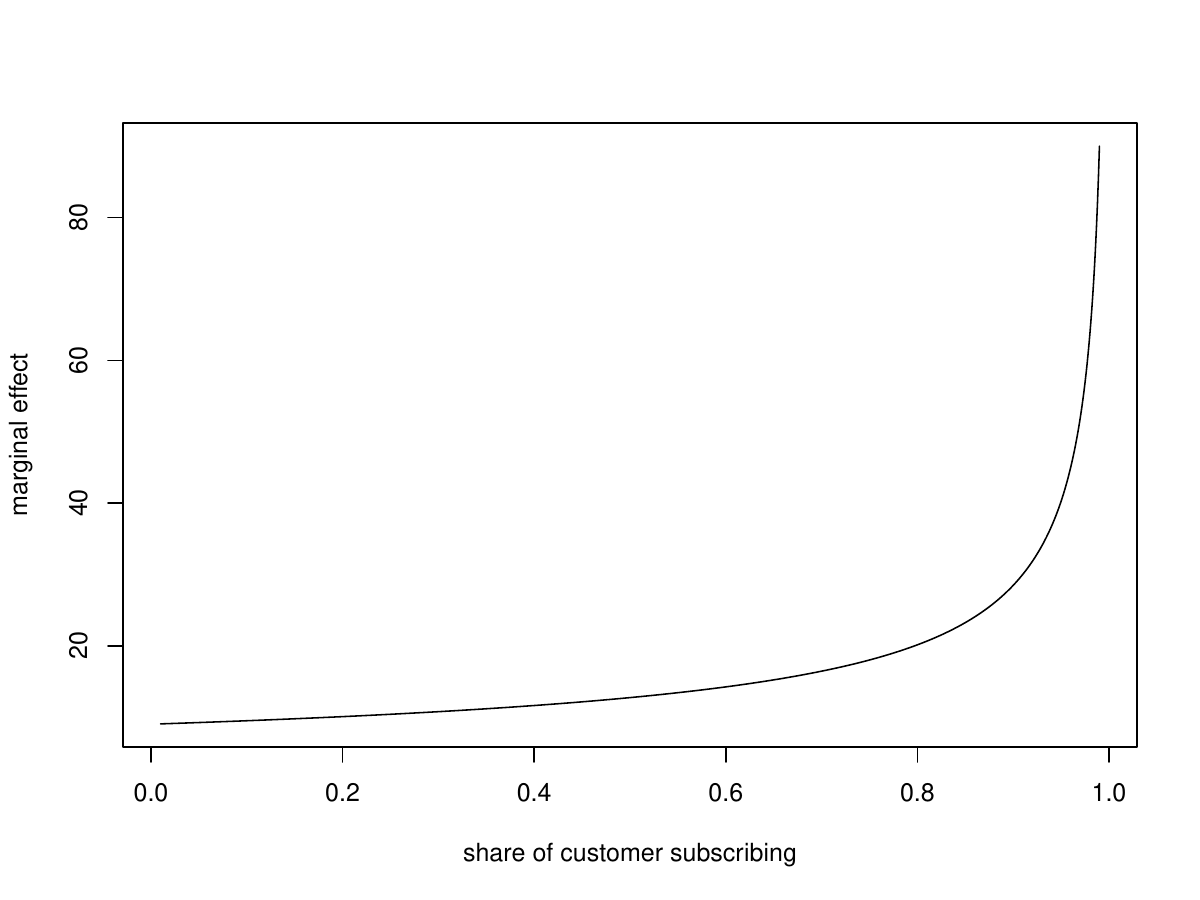}
    \caption{Relation between the percentage subscribing $\beta$ and the corresponding marginal of the expected profit.}
    \label{fig:beta_marginal}
\end{figure}

Figure~\ref{fig:beta_marginal} gives the marginal effect of an additional customer subscribing depending on the share of customers already subscribed $\beta$. As the relative reduction in costs of uncertainty if an additional customer subscribes is stronger when the share of remaining customers ordering at random is small, we obtain a convex relation between $\beta$ and the marginal effect. Thus, if a retailer is able to offer prices sequentially this result should be considered. In particular, this might lead to a situation where it is beneficial to offer discounts that exceed the marginal benefit of first customers subscribing in order to gain the higher benefit from customers ordering at a later stage (or offering subscriptions at all even if it is not beneficial when only a small number of customers subscribe) as also revealed in Implication~4.

\subsubsection*{Summary}
Summarising our results obtained in this section, we find that (1) a subscription offer is particularly beneficial in case the size of the customer base is relatively small and (2) the probability that a single customer buys the SKU is also small. However, we find (3) that there are cases where subscription is not beneficial at all, specifically if supply costs are relatively high compared to the selling price or the buying probability of customers is already high. Regarding the marginal effect of an additional customer subscribing, we (4) find that this effect increases in the number of customers already subscribed. This leads to particularly interesting cases where retailers may offer discounted prices (or subscriptions at all) that reduce the expected profit if only a few customers would agree but increase the expected benefit for larger shares subscribing. These analyses form the basis to gain insights into the optimal price discount offered in case a customer agrees to the subscription offer.

\subsubsection{Deriving the Optimal Price Discount in Case of Subscription}
The retailer is tasked with finding the optimal discount $\tau^*$ that maximises expected profit. As we assume the selling price $p$ as well as supply costs $c$ to be fixed, the retailer is limited to controlling the expected profit by adjusting this price discount in case of a subscription offer. Thus, we aim to derive the effect of the discount on the expected profit. In addition, we have to take into account that this price is likely to affect the share of customers agreeing to the subscription $\beta$. We assume that a higher discount increases the share of customers agreeing, that is, $\frac{\partial \beta}{\partial \tau} >0$.

For the first analysis, we assume that offered prices are not customer-specific, i.e.\ each customer receives the same discount when accepting the subscription offer. The effect of the discount on the expected profit can be obtained by considering the derivative of the expected profit with respect to $\tau$. This effect is given as:
\begin{equation}
    \frac{\partial E(Z_{\text{sub}})}{\partial \tau} = -n\beta + \frac{\partial \beta}{\partial \tau} \left( \delta n + \frac{\gamma}{2} \sqrt{\frac{n \pi (1-\pi)}{1-\beta}} \right) = -n\beta + \frac{\partial E(Z_{\text{sub}})}{\partial \beta} \frac{\partial \beta}{\partial \tau}
    \label{eq:tau_star}
\end{equation}

With an increase in the discounted price, i.e.\ a smaller discount, fewer customers will accept the subscription offer but order at random. Thus, the value of ADI decreases. At the same time, the number of expected customers buying the SKU also decreases. On the other hand, more customers (those ordering at random) pay the full price $p$. These contradicting effects lead to case-specific signs of the marginal effect, particularly depending on the relation between the discount $\tau$ and the share of customers accepting the subscription offer for this discount $\beta$. For the optimal profit $\tau^*$ holds: $\frac{\partial E(Z_{\text{sub}})}{\partial \tau} = 0 \Leftrightarrow n\beta = \frac{\partial E(Z_{\text{sub}})}{\partial \beta} \frac{\partial \beta}{\partial \tau}$.

\section{Computational Experiments}

In the preceding chapter, we explored the influence of uncertainty costs and advanced demand information on the retailer's profit. Additionally, we showcased how subscription offers can mitigate these effects and positively impact profitability. However, our analysis has thus far primarily explored theoretical perspectives. In this chapter, we aim to complement this exploration by incorporating key factors present in practical scenarios. We will conduct a simulation study that encompasses crucial aspects such as an exploration of customer- and SKU-specific characteristics.

\subsection{Setting}
Our simulation-based study considers a retailer that only starts implementing a subscription model. Hence, we start at an initial period without any subscriptions. However, our study considers a given planning horizon covering multiple periods to gain comprehensive insights. Thus, we compare the profit in ongoing periods to the profit obtained in the initial period as a benchmark. By adopting this perspective, we aim to provide valuable insights to retailers at various levels, including those contemplating the implementation of a subscription-based model and those who have already acquired a portion of subscribed customers. We start with a simple model considering homogeneous customers, a single SKU, and only one decision point by the customer to agree or disagree with the subscription offer. This forms the basis for further extensions to the simulation setting.

\subsubsection{Simulation Set Up}
We consider an evaluation period of one year covering 12 months with 4 booking processes in each month (corresponding to one week), totalling 48 evaluation periods. In addition, there is one initial booking process. At the end of this process, the customer has the option to opt for the subscription for the first time. Thus, for inventory management in the initial period, there is no ADI. This period forms a reference for analysing the benefit of subscription offers. We perform $S = 10,000$ simulation runs for each setting to ensure stable results. We aim at determining the discount in case of subscription $\tau^*$ giving the benefit rewarded to agreeing customers that maximises the expected profit of the retailer. All experiments are implemented in Python.

As there is no ADI in the initial period, the optimal order quantity for this period is determined by the number of potential customers $n$, the individual buying probability $\pi$, and the intended strategic service level $\alpha$ as introduced in Section~\ref{sec:customer_demand}. The expected profit is given by Equation~\ref{eq:profit_base}. We now introduce the option of subscription. We begin by considering the case that a customer can only decide in this initial period whether to agree or disagree with the subscription offer. Each customer who orders the SKU in this period will receive such an offer. In case the customer agrees, there will be a delivery in each of the evaluation periods associated with the discount $\tau$. Otherwise, the customer will remain the individual buying probability $\pi$ for the evaluation periods. Initially, we assume that the retailer receives all orders at the same point in time. Thus, the benefit rewarded to customers agreeing to the subscription-based model cannot be personalised but needs to be equal for all customers with the same buying probability. As we assume that the decision on agreeing to the subscription takes place in the first period only, in the basic setting, the expected profit is equal for all evaluation periods as given in Equation~(\ref{eq:profit_sub}).

Equation~(\ref{eq:tau_star}) yields the condition for the optimal discount $\tau^*$. However, in general, there is no closed-form solution for this expression. Thus, we rely on a numerical search based on the average profit over a set of simulation runs $S$ within our simulation scenario to determine the discount $\tau^*$. In particular, $z_{s,t}(\tau)$ denotes the (realised) profit for period $t \in T$ and simulation run $s \in S$ based on the realisations of demand in each period and potentially the acceptance of a subscription offer in the first period for a given discount $\tau$. As we can consider all evaluation periods equivalently in this basic scenario, we can calculate the average profit $\bar{z} = \sum\limits_{s \in S} \sum\limits_{t \in T} z_{s,t}$ as a proxy for the expected profit which we want to maximise. Assuming that subscribed customers are served first, the profit for a single evaluation period of a simulation run is given by the following equation:
\begin{equation*}
    z_{s,t} = (p-\tau) \cdot n^{sub}_s + p \cdot (\min\{x_{s,t}, q_{s,t}\} - n^{sub}_s) - c \cdot q_{s,t}
\end{equation*}
where $n^{sub}_s$ corresponds to the number of customers subscribed in this simulation run; $x_{s,t}$ and $q_{s,t}$ give the demand and order quantity in period $t$ and simulation run $s$, respectively. 

\subsubsection{Reaction of Customers to Subscription Offers}
\label{sec:customer_reaction}

We assume that each consumer has a characteristic $\lambda \in (0,1)$ which corresponds to the individual popularity of a subscription-based service, capturing, e.g.\ effects of convenience and a lower risk of shortage \citep{gallego2001integrating}. As introduced above, we denote the discount offered to the customer in case of accepting the subscription by $\tau$. As we fixed the selling price to $p = 1$, obviously $\tau \in [0,1]$. We further assume that the acceptance probability depends on the individual buying probability, i.e.\ it is more likely that a customer agrees to the subscription offer if he/she orders an SKU often anyway. We assume that a customer won't agree to the subscription offer if he/she either (1) does not buy the SKU (i.e.\ the buying probability is 0), (2) does not like subscriptions at all, or (3) won't receive any benefit for agreeing. Thus, we rely on a Cobb-Douglas function with three input factors and equal weights to approximate the probability $\eta$ of accepting the offer:
\begin{equation}
    \eta = (\tau \cdot \pi \cdot \lambda)^{1/3}.
\end{equation}
This relation ensures probabilities $\eta \in [0,1]$. As only customers who order the SKU during the initial booking process receive such an offer, we denote the ex-ante probability, i.e.\ before the start of the initial booking process, that a customer subscribes by $\beta = \pi \cdot \eta$. This allows us to explicitly formulate the expected profit with the functional relation $\beta(\tau)$ for our simulation experiments. 

\begin{figure}[htp!]
    \centering
    \includegraphics[width=0.6\textwidth]{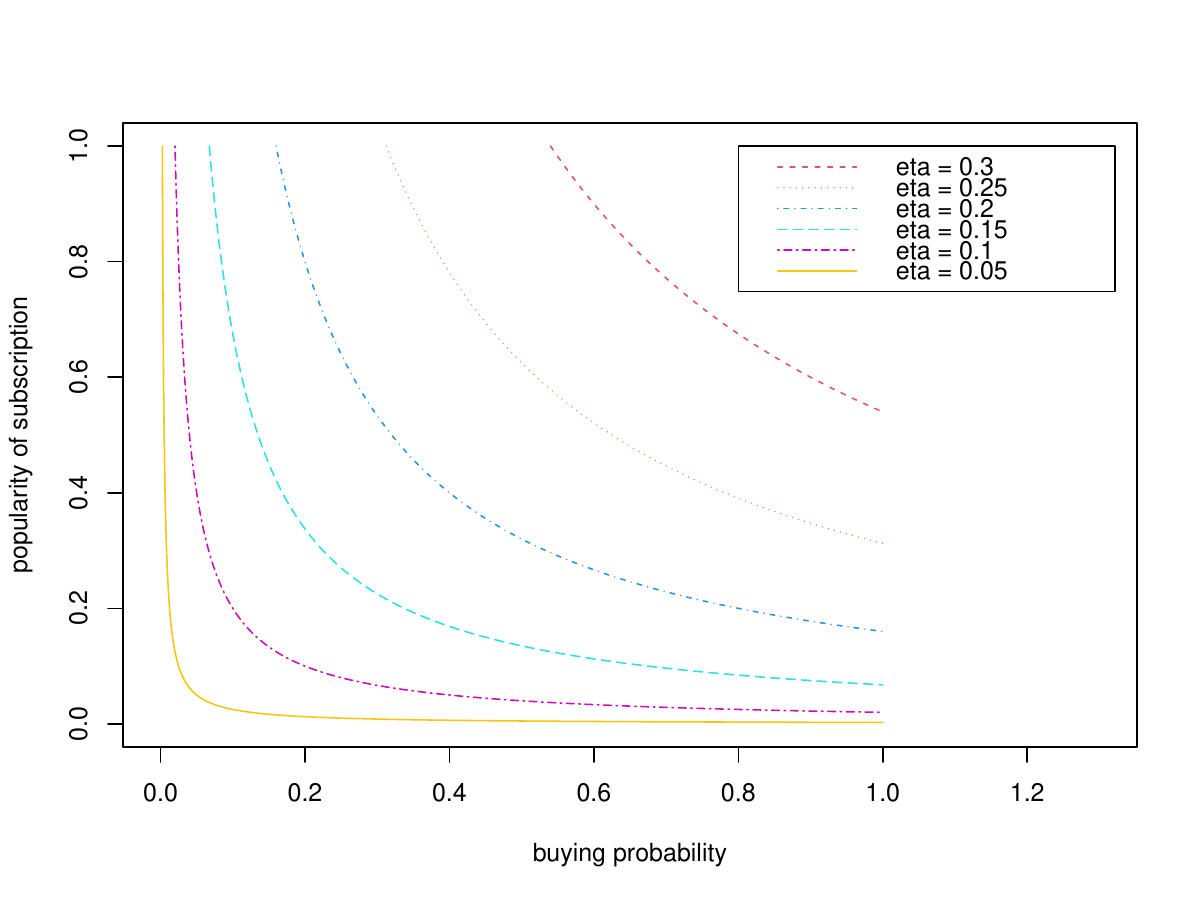}
    \caption{Combinations of the individual buying probability $\pi$ and the popularity of subscription $\lambda$ leading to the same probability to accept a subscription offer in case of a fixed price discount of $\tau = 7.5\%$.}
    \label{fig:prob_sub}
\end{figure}

Figure~\ref{fig:prob_sub} displays isoquants, i.e.\ combinations of the individual buying probability $\pi$ and the popularity of subscription $\lambda$ leading to the same probability to accept a subscription offer, for a discount of 7.5\%. The yellow line corresponds to $\eta = 5\%$ acceptance probability, and the red line to $\eta = 30\%$. The other lines are drawn in steps of 5 percentage points in the acceptance probability. We find that such high acceptance probabilities occur only if both the popularity of subscriptions and the individual buying probability are relatively high. On the other hand, an acceptance probability of 5\% is given even when both probabilities are relatively small. In case of a higher discount, we would observe a shift to the bottom left of all isoquants, i.e.\ lower individual buying probabilities and lower popularity of subscriptions are sufficient to obtain the same acceptance probability.

\subsection{Basic Example}
\begin{figure}[htp!]
    \centering
    \includegraphics[width=0.6\textwidth]{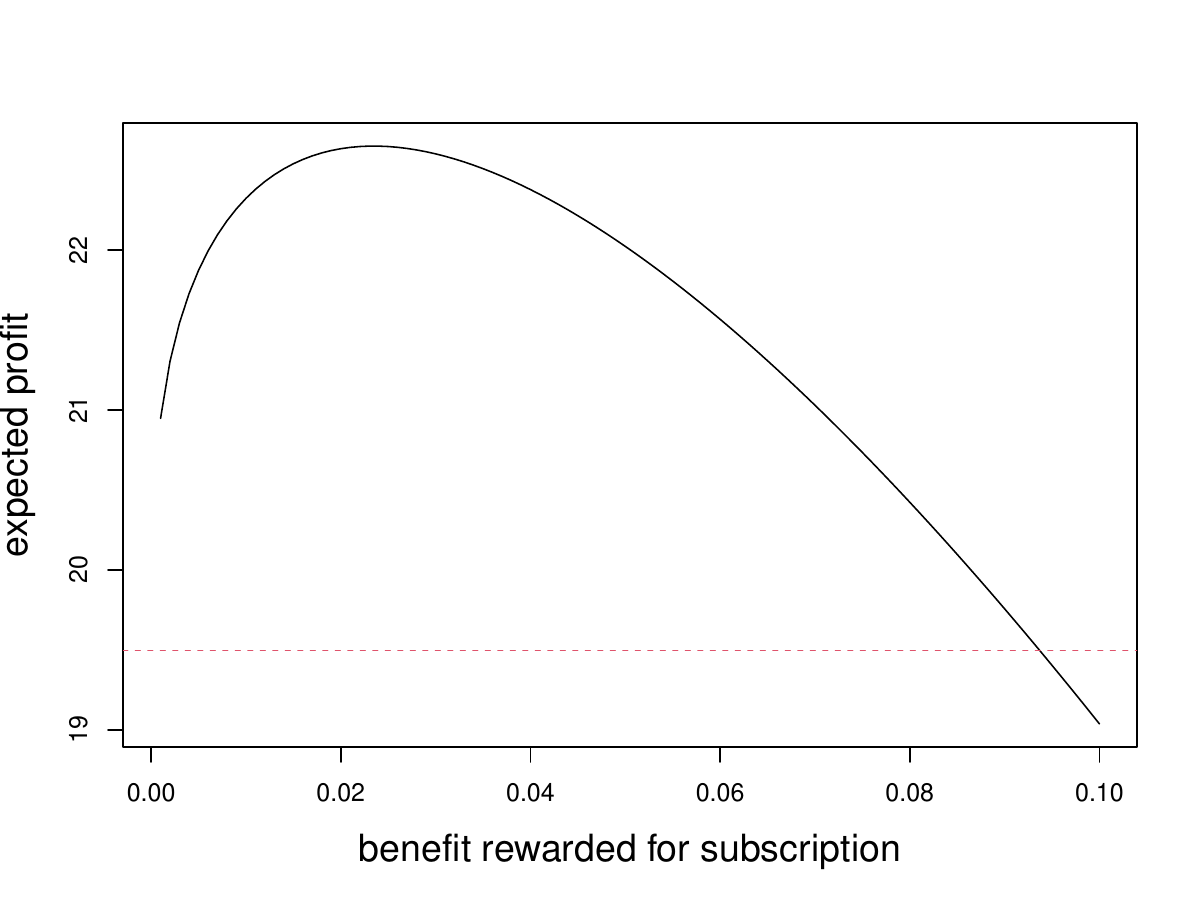}
    \caption{Relation between the benefit rewarded to subscribing customers $\tau$ and the expected profit in optimum. The red dotted line indicates the profit without subscription offers}
    \label{fig:benefit_profit}
\end{figure}

Similar to the numerical examples in the section above, we initially fix the number of potential customers to $n = 500$, the individual buying probability to $\pi = 0.5$, the selling price to $p = 1$, supply costs to $c = 0.85$, and the strategic service level to $\alpha = 0.97$. We assume a popularity of subscription $\lambda = 0.5$. We start by analytically deriving the profit. In the situation without subscriptions, i.e.\ the initial booking process, we can obtain the expected profit from Equation~\ref{eq:profit_base} as $E(Z_1) = 19.50$ (see Table~\ref{tab:uncertainty}). We now offer each customer ordering in the initial booking process the option to subscribe. Figure~\ref{fig:benefit_profit} gives the relation between the rewarded benefit $\tau$ and the expected profit under the subscription model. We find an optimal discount of $\tau^* = 2.3\%$ associated with an expected revenue of $22.65$ according to Eqaution~\ref{eq:profit_sub}. The expected profit is 16.2\% higher compared to the expected profit in the initial period (without subscription offers) indicated by the red dotted line in Figure~\ref{fig:benefit_profit}.

Applying our simulation setting to the same scenario, we find an average profit over all simulation runs in the initial booking process of $\bar{Z}_1^{\text{sim}} = 19.54$ and with subscriptions of $\bar{Z}^{\text{sim}}_{\text{sub}} = 22.68$ which is very close to the analytical results. Thus, we can confirm the adequacy of our simulation analysis as well as the positive effect of introducing a subscription-based service in this setting. We want to consider our results in more detail. The expected share of customers subscribing in our analytical model is given by $\beta = 8.90\%$. This is similar to the average share of customers subscribing across our simulation runs $\beta^{\text{sim}}$. The difference in the expected profit can be derived by considering two different effects: the expected profit for the SKUs sold (part I of the objective function) is affected by the expected demand and the share of customers subscribing. According to Equation~\ref{eq:tau} this effect is positive in our case as $\tau = 0.023 < 0.075 = \delta$. Specifically, we can calculate the increase in the expected profit driven by these parts of the objective function as $I^{\Delta} = n \cdot \beta \cdot \delta = 2.33$. In addition, the reduction in uncertainty affects the expected profit. This effect is given as $II^{\Delta} = \gamma \cdot \sqrt{n \cdot \pi \cdot (1-\pi)} \cdot (1-\sqrt{1-\beta}) = 0.82$. This accumulates to a total increase in the expected profit of $I^{\Delta} + II^{\Delta} = 2.33 + 0.82 = 3.15$. Thus, we can summarise that under the optimal price discount for customers subscribing the expected profit increases due to two effects: (1) the increase in the expected number of customers demanding the SKU and (2) the reduction in costs of uncertainty.

\subsection{Impact of the Model Parameters}
After we have considered a basic scenario with fixed parameters in detail, we want to gain insights into how the different model parameters affect the optimal benefit rewarded to customers agreeing to the subscription as well as the potential increase in the expected profit of the retailer when introducing subscription offers. While we assume the selling price to be fixed to $p = 1$ and the service level to be strategically given as $\alpha = 0.97$, we separate the following analyses into (1) SKU-specific characteristics, namely the customer base $n$ as well as supply costs $c$ and (2) customer-specific characteristics, namely the individual buying probability $\pi$ and the popularity of subscription $\lambda$. 

\subsubsection*{Product-Specific Characteristics: size of the Customer Base and Supply Costs}
As we showed in Section~\ref{sec:impact_uncertainty}, the importance of uncertainty for the expected profit reduces with an increase in the customer base $n$. Thus, we are interested in the effects in case of a subscription-based model. Therefore, we again consider different values of the customer base in the interval $n \in [150, 1000]$ and determine the discount $\tau^*$ for each scenario, where we refer to parameters for the other variables as introduced above in the basic example. This allows us to analyse the effect of the customer base $n$ on a possible change in the discount $\tau$, the corresponding share of customers subscribing $\beta$, as well as the relative increase in the expected profit when introducing subscription offers. We perform the same analyses for supply costs $c \in (0, 0.9]$ as we showed in Figure~\ref{fig:c_effect} that costs of uncertainty are highly dependent on the supply costs. In all cases, we consider this cost relative to the fixed selling price of $p = 1$.

\begin{figure}[ht]
    \centering
    \subfigure[customer base vs.\ optimal discount]{\includegraphics[width=0.42\textwidth]{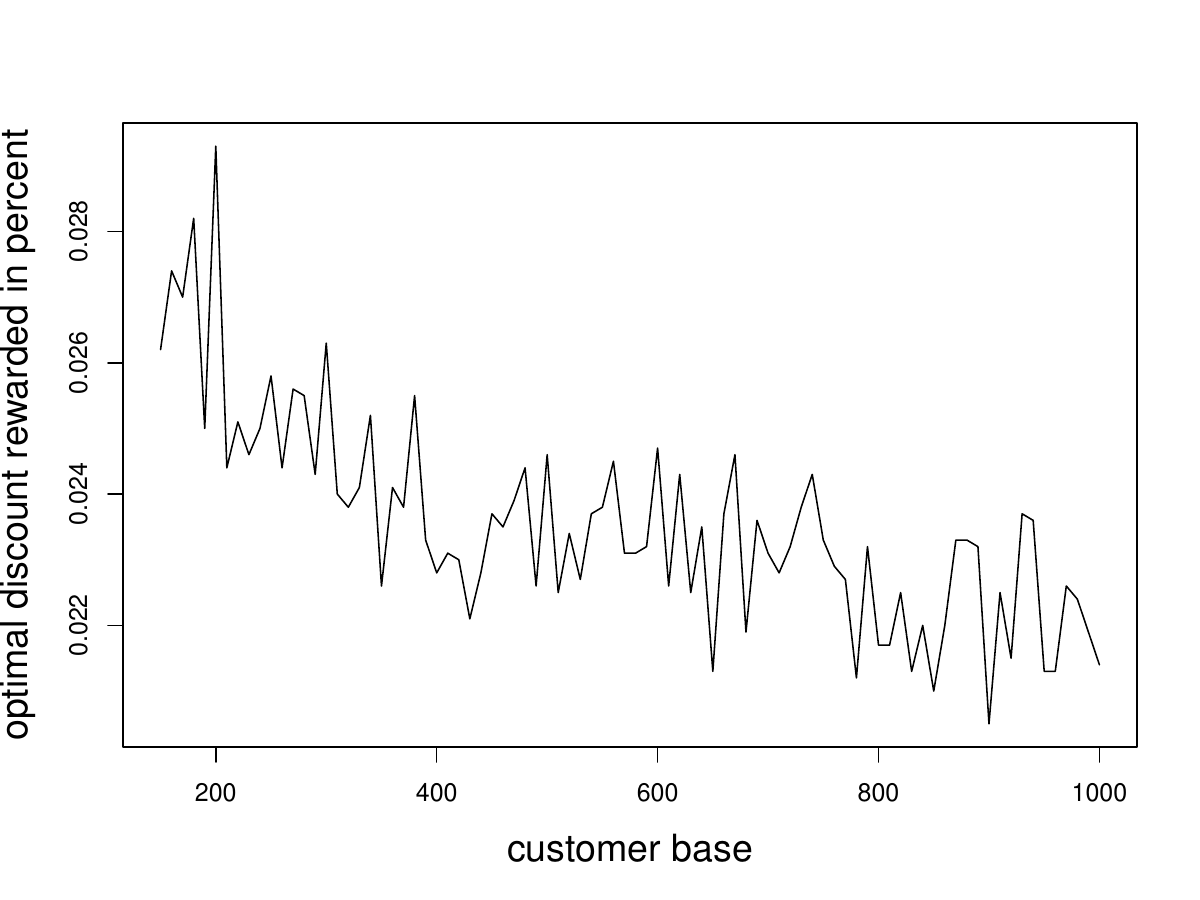}}
    \subfigure[customer base vs.\ share subscribing]{\includegraphics[width=0.42\textwidth]{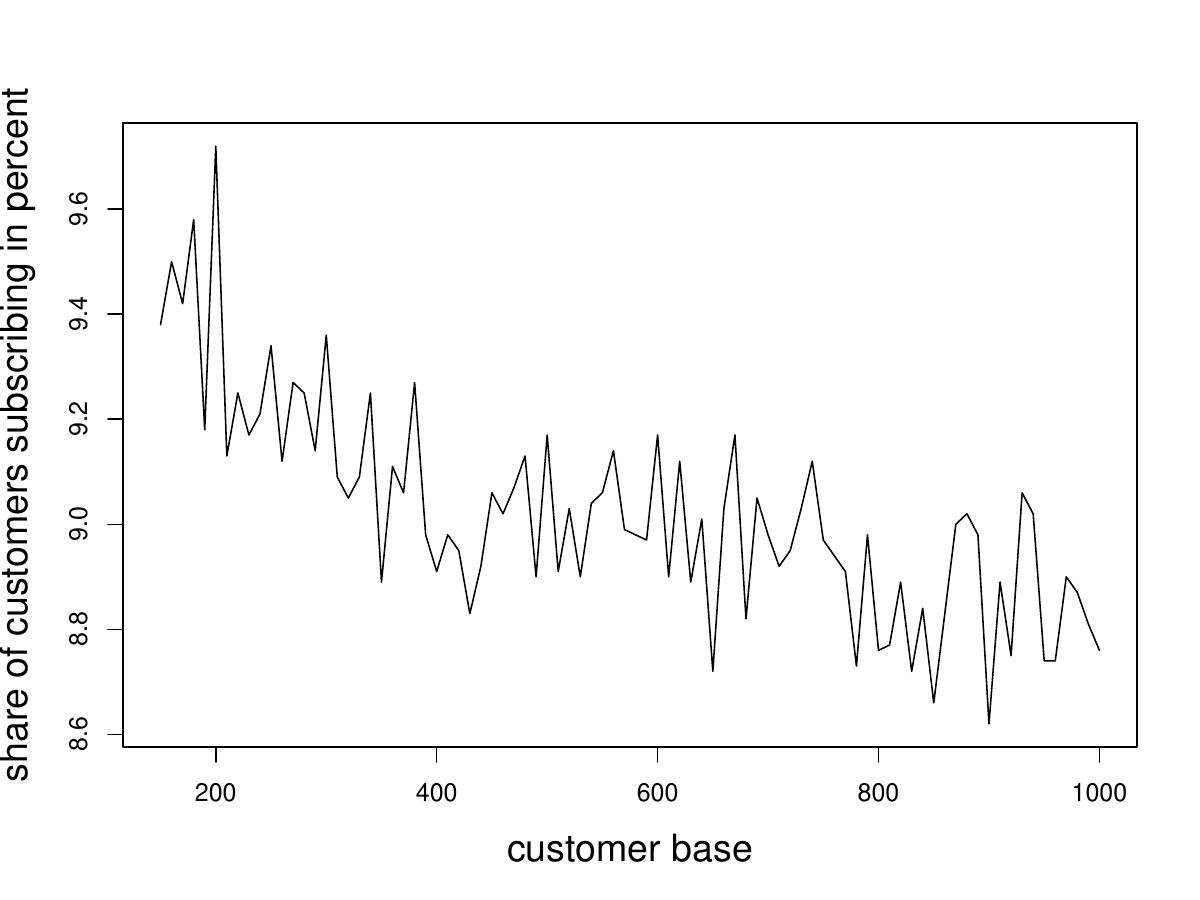}}
    \subfigure[customer base vs.\ benefit of subscription]{\includegraphics[width=0.42\textwidth]{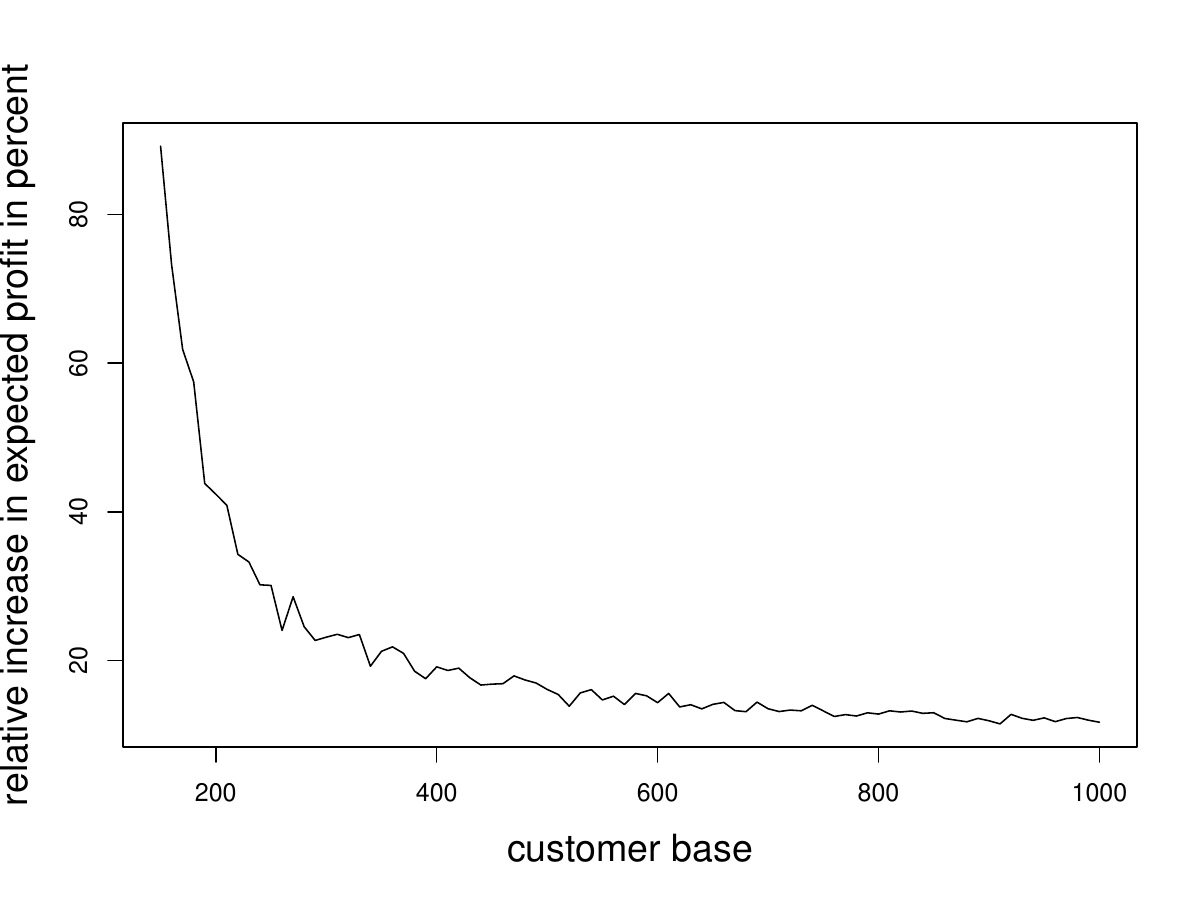}}
    \caption{Relation between the size of the customer base $n$ and (a) the optimal price discount $\tau^*$, (b) the share of customers subscribing $\beta$, and (c) the increase in the expected profit with subscriptions.}
    \label{fig:n_sub}
\end{figure}

Figure~\ref{fig:n_sub} gives the relation between the size of the customer base $n$ on the x-axis and (a) the optimal benefit rewarded to subscribing customers $\tau^*$ in optimum, (b) the share of customers subscribing $\beta$, and (c) the relative increase in the expected profit when introducing the subscription offer and referring to the discount derived in (a). We find that the optimal discount slightly decreases with an increase in the customer base from about 3\% for $n \approx 200$ to about 2\% for $n \approx 1000$. This aligns with our previous result that the benefit of ADI is smaller for a larger customer base. Thus, the benefit of subscription offers is also smaller and induces retailers to reward a smaller benefit to customers subscribing; the share of customers subscribing consequently also reduces. These two aligned effects lead to a strong decline in the relative increase of the expected profit. While this increase lies above 20\% for $n < 300$ it declines to still more than 10\% when $n$ approaches $1000$. However, our results again confirm the positive impact of implementing subscription offers for the retailer, particularly beneficial for smaller customer bases.

\begin{figure}[ht]
    \centering
    \subfigure[supply costs vs.\ optimal discount]{\includegraphics[width=0.42\textwidth]{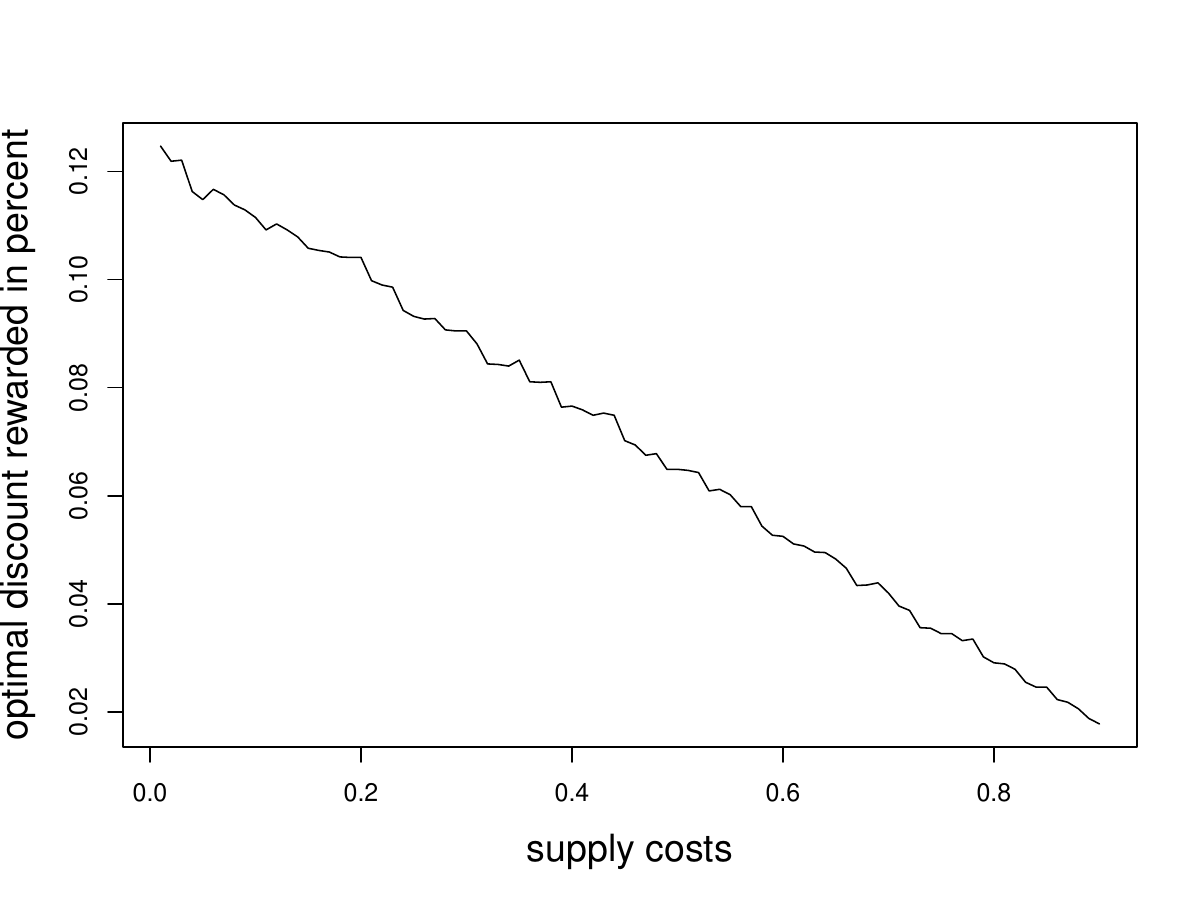}}
    \subfigure[supply costs vs.\ share subscribing]{\includegraphics[width=0.42\textwidth]{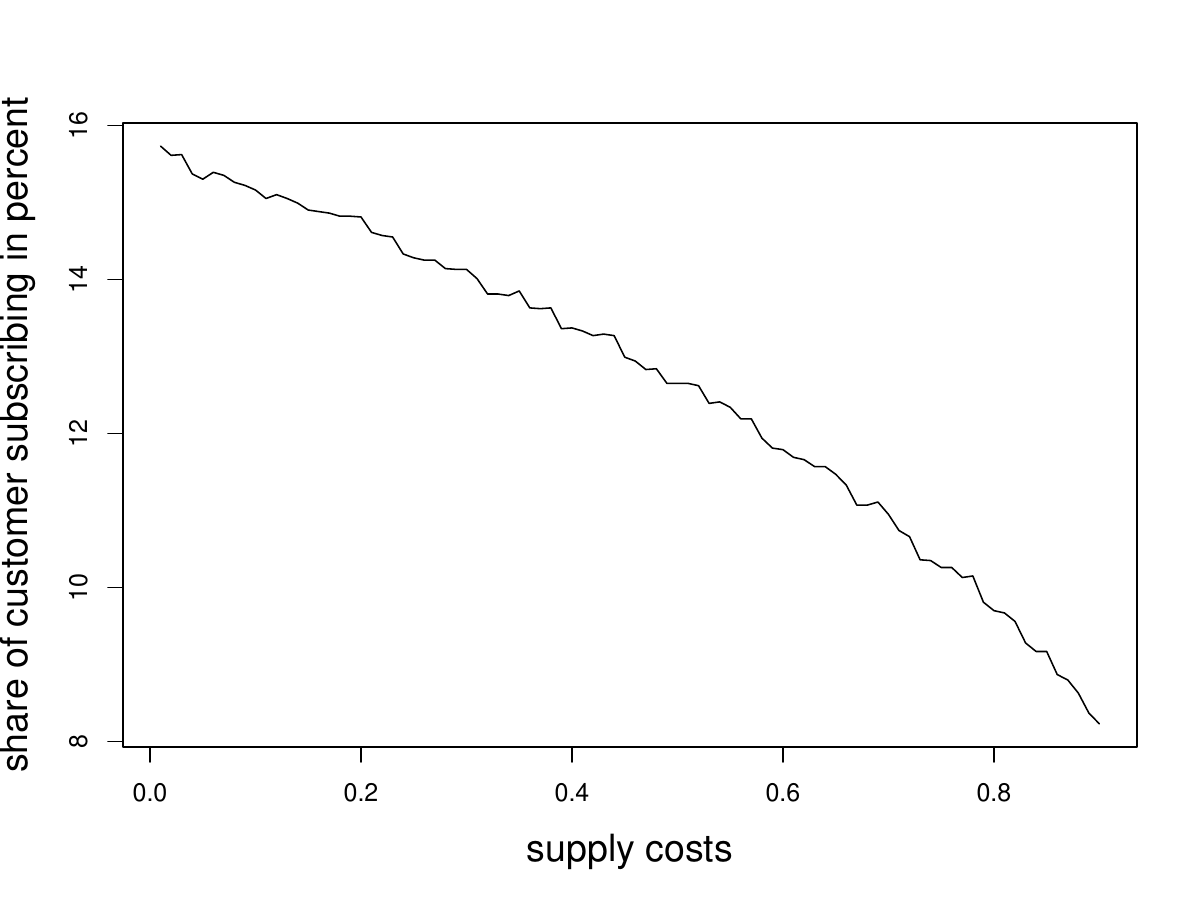}}
    \subfigure[supply costs vs.\ benefit of subscription]{\includegraphics[width=0.42\textwidth]{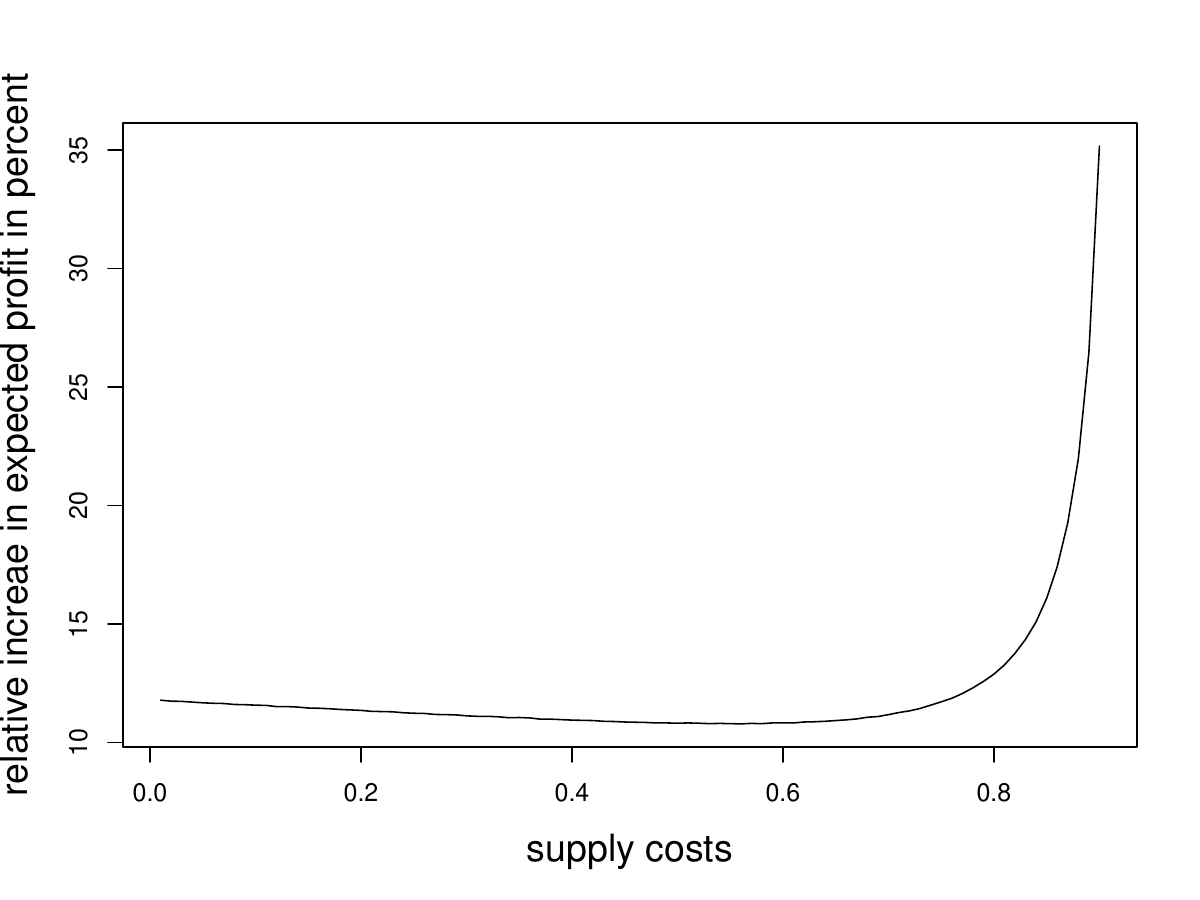}}
    \caption{Relation between supply costs $c$ and (a) the optimal price discount $\tau^*$, (b) the share of customers subscribing, and (c) the increase in the expected profit.}
    \label{fig:c_sub}
\end{figure}

Figure~\ref{fig:c_sub} provides the same statistics for different values of supply costs $c$. First, we find a nearly linear decrease in the benefit rewarded to customers subscribing $\tau^*$ from more than 10\% if supply costs are fairly small to about 2\% for $c \approx 0.9$ (see Figure~\ref{fig:c_sub} (a)). Thus, the retailer has a higher incentive to offer a higher benefit in case the supply costs are low, i.e.\ the short-term margin of the retailer is large. The assumed relation between the price discount and the share of customers subscribing translates this linear decrease into a concave decrease in the share of customers subscribing. The interplay between these two effects together with a smaller absolute difference between the expected profit without and with subscription offer (see Figure~\ref{fig:c_effect}) leads to the situation that the relative increase in the expected profit is lowest for $c \approx 0.55$. Afterwards, implementing a subscription offer yields a much higher expected profit relative to the profit without subscriptions with an increase of more than one-third for high supply costs of $c = 0.9$.

\subsubsection*{Customer-Specific Characteristics: Buying Probability and Popularity of Subscription}
As discussed in Section~\ref{sec:customer_reaction}, customers vary in their individual buying probability as well as the individual popularity of subscriptions. To account for this effect, we allow for heterogeneity across customers. To this end, we consider five different settings for the popularity of subscription, namely \textit{very high} (95\%), \textit{high} (75\%), \textit{medium} (50\%), \textit{low} (25\%), and \textit{very low} (5\%). For the individual buying probability, we consider the same settings except for \textit{very low} as this would lead to a negative expected profit. For each combination, we calculate the discount $\tau^*$, the average share of customers subscribing $\beta$, and the relative increase in the expected profit $\Delta Z$ when introducing a subscription service based on our simulation setting. This allows the retailer for price discrimination in terms of the benefit rewarded to each customer segment. However, such discrimination requires the company to know about the type of customer regarding the individual buying probability and the popularity of subscriptions. Otherwise, if the retailer only knows the proportion of each group, all customers need to receive the same offer. In this case, at least we are able to derive the effect of different proportions of each group on the price discount and the expected profit.
      
\begin{table}[t]
    \centering
    \scalebox{0.6}{
    \begin{tabular}{r|ccccc}
      & \multicolumn{5}{c}{Popularity of subscription} \\ 
      Individual buying probability & Very low: $\lambda = 0.05$ & Low: $\lambda = 0.25$ & Medium: $\lambda = 0.5$ & High: $\lambda = 0.75$ & Very high: $\lambda = 0.95$ \\
        \hline                      
         & $\tau^* = 3.2\%$ & $\tau^* = 3.1\%$ & $\tau^* = 3.3\%$ & $\tau^* = 3.1\%$ & $\tau^* = 3.3\%$ \\
        Low: $\pi = 0.25$ & $\beta = 1.84\%$ & $\beta = 3.12\%$ & $\beta = 3.99\%$ & $\beta = 4.51\%$ & $\beta = 4.95\%$ \\
         & $\Delta Z = 24.42\%$ & $\Delta Z = 43.77\%$ & $\Delta Z = 55.79\%$ & $\Delta Z = 64.02\%$ & $\Delta Z = 69.62\%$ \\
        \hline   
         & $\tau^* = 2.3\%$ & $\tau^* = 2.3\%$ & $\tau^* = 2.5\%$ & $\tau^* = 2.3\%$ & $\tau^* = 2.3\%$ \\
        Medium: $\pi = 0.5$ & $\beta = 4.20\%$ & $\beta = 7.11\%$ & $\beta = 9.17\%$ & $\beta = 10.29\%$ & $\beta = 11.06\%$ \\
         & $\Delta Z = 7.42\%$ & $\Delta Z = 12.76\%$ & $\Delta Z = 16.09\%$ & $\Delta Z = 18.44\%$ & $\Delta Z = 19.91\%$ \\        
        \hline    
         & $\tau^* = 1.4\%$ & $\tau^* = 1.4\%$ & $\tau^* = 1.4\%$ & $\tau^* = 1.3\%$ & $\tau^* = 1.3\%$ \\
        High: $\pi = 0.75$ & $\beta = 6.01\%$ & $\beta = 10.29\%$ & $\beta = 13.14\%$ & $\beta = 14.74\%$ & $\beta = 15.72\%$ \\
         & $\Delta Z = 2.67\%$ & $\Delta Z = 4.82\%$ & $\Delta Z = 6.15\%$ & $\Delta Z = 7.12\%$ & $\Delta Z = 7.71\%$ \\
         \hline         
         & $\tau^* = 0.4\%$ & $\tau^* = 0.3\%$ & $\tau^* = 0.4\%$ & $\tau^* = 0.4\%$ & $\tau^* = 0.3\%$ \\
        Vera high: $\pi = 0.95$ & $\beta = 5.39\%$ & $\beta = 8.89\%$ & $\beta = 12.11\%$ & $\beta = 13.35\%$ & $\beta = 13.61\%$ \\
         & $\Delta Z = 0.41\%$ & $\Delta Z = 0.74\%$ & $\Delta Z = 0.97\%$ & $\Delta Z = 1.12\%$ & $\Delta Z = 1.21\%$ \\
    \end{tabular}}
    \caption{Results on the optimal benefit rewarded $\tau^*$, the expected share of customers agreeing to the subscription offer $\beta$, and the relative increase in the profit $\Delta Z$ for different levels of the individual buying probability $\pi$ and the popularity of subscription $\lambda$ for the basic parameter setting.}
    \label{tab:buying_popularity}
\end{table}

Table~\ref{tab:buying_popularity} gives comprehensive results regarding the effect of the individual buying probability $\pi$ and the popularity of subscription $\lambda$ on the optimal benefit rewarded $\tau^*$, the expected share of customers agreeing to the subscription offer $\beta$, and the relative increase in the profit $\Delta Z$. We find that the popularity of subscriptions does not affect the benefit rewarded to customers when agreeing on the subscription offer. At the same time, the retailer should reward a higher benefit if the individual buying probability is low. This is plausible as it is more beneficial for retailers if a customer who does not order an SKU regularly commits to the subscription offer and agrees to receive this SKU in every demand period as this leads to a stronger increase in the expected revenue. Obviously, the average share of customers agreeing to the subscription offer $\beta$ increases in the popularity of subscription $\lambda$. Regarding the individual buying probability $\pi$ two contradicting effects co-exist: (1) a higher buying probability increases the probability of agreeing to a subscription offer for a given popularity and benefit; (2) the benefit rewarded decreases in the buying probability.

This leads to the situation that the share of customers agreeing to the subscription offer increases until the high level of individual buying probability ($\pi = 0.75$) and decreases for the very high level ($\pi = 0.95$). We find the highest share $\beta = 15.72\%$ for the high level of buying probability and the very high level of popularity. From the perspective of the retailer, customers with a very high popularity and a very low individual buying probability are most beneficial for introducing the subscription model. In this case, we find an increase in the expected profit when introducing the subscription model of nearly 70\%. This is mainly driven by the increase in the expected number of customers ordering the SKU if such a customer agrees to regular orders, even if the share of customers agreeing to the subscription offer is relatively small (below 5\%). On the other hand, for SKUs with a very high buying probability the increase in the expected profit $\Delta Z$ is very small (below 1\% even for the medium level of popularity).

In conclusion, retailers should focus on SKUs with a small customer base $n$ and either high or very small margins induced by the supply costs $c$. Offering the option for subscription is most beneficial for customers with a small buying probability. This holds particularly for customers with high popularity for subscriptions, even if the share of customers agreeing to the offer is relatively small.

\subsection{Outlook}

Our simulation setting allows for various analyses regarding the impact of implementing subscription offers on the expected profit of the retailer. So far, we considered products individually and segmented customers according to their individual buying probability and the popularity of subscriptions. At the same time, we account for heterogeneity across products by different sizes of the customer base and supply costs relative to a normalised selling price. However, we made restrictive assumptions regarding the arrival of customers during the booking process. In particular, we assume that all customers arrive at the same time; thus, receiving the same subscription offer in terms of the benefit rewarded. In fact, in retail practice, customer orders arrive sequentially. This allows retailers to make individual subscription offers to different customers arriving at different points in time. This is particularly relevant in light of our result that the marginal effect of an additional customer subscribing does depend on the share of customers already subscribing. This extension can be addressed by modelling the problem by a sequential decision process.

In our simulation model, we consider one year with a booking process every week. In the analyses already performed we allow for a decision whether to subscribe or not in the initial booking process only. This raises the possibility for further extensions: first, the retailer can place an offer for agreeing to a subscription-based service at the end of each booking process instead of only in the first phase. For example, if a customer disagrees with the offer in the initial phase, he/she might be willing to accept the offer at a later stage. In particular, the optimal price discount is affected by the share of customers already agreed to subscriptions. Thus, the benefit rewarded might be higher within the course of the year, thus, attracting further customers to agree if an offer is placed at that time. Second, our introductory example illustrated in Figure~\ref{fig:customer_orders} reveals different buying patterns. Specifically, some products are ordered, e.g.\ on a biweekly basis or once a month. Thus, a customer might not be willing to agree to an offer that leads to a weekly delivery but to another pattern even if the discount might be smaller. Another extension could be directed to the length of the subscription offer. While we consider a time horizon of one year, some customers might have a higher willingness to agree to a subscription offer that lasts for, e.g.\ three months only. To summarise, our approach and simulation setting form the basis for ongoing research and interesting extensions that meet the characteristics of e-grocery retailing and its customers even more accurately.

\section{Conclusion}

In this paper, we have explored the implementation of subscription offers within the e-grocery sector to address inventory challenges. As subscription offers typically involve price reductions, we have outlined a three-step procedure ensuring that offering price reductions positively impacts profitability. Our approach involves first calculating the expected planning costs of uncertainty, assessing the value of advanced demand information, and then determining the appropriate level of price reduction. Throughout our analysis, we have considered varying purchase probabilities for different products, as well as the diverse costs incurred by retailers when inaccurately anticipating inventory levels. Additionally, we recognise that not all customers will be willing to subscribe to offers. Our findings reveal several insights:
\begin{itemize}
\item Uncertainty costs significantly impact the retailer's overall profit, particularly in an e-grocery context with high service level targets and narrow profit margins. This effect is exacerbated for smaller customer bases, which is pertinent for new businesses entering the e-grocery market. Products with around 50\% buying probabilities experience particularly high uncertainty costs.
\item Collecting advanced demand information can mitigate uncertainty costs, leading to potential profit increases of more than 60\%, especially for products with low buying probabilities. Higher participation rates among customers in sharing their purchase intentions lead to greater relative profit gains, highlighting the importance of customer engagement.
\item Subscription offers are most beneficial for smaller customer bases and products with low buying probabilities. However, there are cases where subscription offers may not be advantageous, such as when supply costs are high relative to selling prices or when customer buying probabilities are already high. Our three-step approach enables retailers to discern precisely where implementing subscriptions enhances profitability and where it may not be beneficial.
\end{itemize}

While the initial portion of our paper delved into the theoretical aspects of subscriptions, we augmented this perspective with a simulation study to gain insights into practical implications. Our experiments encompassed various product-specific characteristics, including the number of customers purchasing a specific product and the associated supply costs. Additionally, we considered customer-specific attributes such as buying probabilities and the likelihood of agreeing to subscription offers.

Moving forward, our aim is to translate our findings into actionable policies that can be readily implemented by retailers. Although our study primarily focused on assessing the value of subscription offers for individual products and customer segments, we recognise the potential confusion that could arise from offering multiple subscription options at various prices to customers with diverse basket sizes. Therefore, we intend to explore strategies for crafting customer-friendly offers while maintaining overall profitability. Furthermore, we seek to investigate the effects of long-term subscriptions, such as extending subscription offers over a period aligned with the customer's willingness to commit to such arrangements. By doing so, we aim to gain a deeper understanding of the long-term implications of subscription models on both retailer profitability and customer satisfaction.

\bibliographystyle{apalike} 
\bibliography{library}

\end{spacing}
\end{document}